\documentclass[preprint]{imsart}

\RequirePackage[OT1]{fontenc}
\RequirePackage[numbers]{natbib}

\usepackage{amsthm,amsmath}
\usepackage{amsbsy}

\input xy
\xyoption{all}

\usepackage{algorithm}
\usepackage{algpseudocode}

\usepackage{epsf}
\usepackage{epsfig}

\newcommand{\beginsupplement}{%
        \setcounter{table}{0}
        \renewcommand{\thetable}{S\arabic{table}}%
        \setcounter{figure}{0}
        \renewcommand{\thefigure}{S\arabic{figure}}%
     }

\startlocaldefs

\newcommand{\ind}{1\!\!1}
\newcommand{\bfU}{\boldsymbol{U}}
\newcommand{\bfE}{\boldsymbol{E}}
\newcommand{\bfV}{\boldsymbol{V}}
\newcommand{\bfX}{\boldsymbol{X}}
\newcommand{\bfy}{\boldsymbol{y}}
\newcommand{\bfzero}{\boldsymbol{0}}
\newcommand{\bfone}{\boldsymbol{1}}
\newcommand{\bfSigma}{\boldsymbol{\Sigma}}
\newcommand{\bfPsi}{\boldsymbol{\Psi}}
\newcommand{\bfI}{\boldsymbol{I}}

\newcommand{\bfW}{\boldsymbol{W}}
\newcommand{\bfM}{\boldsymbol{M}}
\endlocaldefs

\begin{document}

\begin{frontmatter}


\title{Learning Disease vs Participant Signatures: a permutation test approach to detect identity confounding in machine learning diagnostic applications}
\runtitle{Detecting identity confounding in diagnostic applications}



\author{\fnms{} \snm{Elias Chaibub Neto$^{\ast, 1}$, Abhishek Pratap$^{1, 2}$, Thanneer M \\ Perumal$^1$, Meghasyam Tummalacherla$^1$, Brian M Bot$^1$, \\ Andrew D Trister$^1$, Stephen H Friend$^1$, \\ Lara Mangravite$^1$, and Larsson Omberg$^1$}\ead[label=e1]{$^\ast$ elias.chaibub.neto@sagebase.org, $^1$ Sage Bionetworks, $^2$ Department of Biomedical Informatics and Medical Education, University of Washington, Seattle}}
\address{\printead{e1}}



\runauthor{Chaibub Neto E. et. al.}

\begin{abstract}
Recently, Saeb \textit{et al} (2017) showed that, in diagnostic machine learning applications, having data of each subject randomly assigned to both training and test sets (record-wise data split) can lead to massive underestimation of the cross-validation prediction error, due to the presence of ``subject identity confounding" caused by the classifier's ability to identify subjects, instead of recognizing disease. To solve this problem, the authors recommended the random assignment of the data of each subject to either the training or the test set (subject-wise data split). The adoption of subject-wise split has been criticized in Little \textit{et al} (2017), on the basis that it can violate assumptions required by cross-validation to consistently estimate generalization error. In particular, adopting subject-wise splitting in heterogeneous data-sets might lead to model under-fitting and larger classification errors. Hence, Little \textit{et al} argue that perhaps the overestimation of prediction errors with subject-wise cross-validation, rather than underestimation with record-wise cross-validation, is the reason for the discrepancies between prediction error estimates generated by the two splitting strategies. In order to shed light on this controversy, we focus on simpler classification performance metrics and develop permutation tests that can detect identity confounding. By focusing on permutation tests, we are able to evaluate the merits of record-wise and subject-wise data splits under more general statistical dependencies and distributional structures of the data, including situations where cross-validation breaks down. We illustrate the application of our tests using synthetic and real data from a Parkinson's disease study.
\end{abstract}





\end{frontmatter}

\section{Introduction}

The data-driven diagnosis of diseases is currently a major focus of applied machine learning research. Under the supervised learning paradigm, a classifier of disease status (i.e., cases versus controls) is built by training the machine learning classifier on example data consisting of input variables (features) and an output variable (case and control labels), and then making predictions about unseen labels based on newly observed feature data.

An important element of this endeavor is to assess the prediction performance of the trained classifier. The key point is to evaluate the performance of the algorithm on a completely separate data-set (test set) other than the one used to train the classifier (training set), so that the algorithm doesn't make correct label predictions just because it ``remembers" the examples it already saw. However, because diagnostic systems based on mobile health applications are able to collect dense longitudinal data on each user, the question then arises on how should one split the data when evaluating the performance of a diagnostic machine learning system\cite{saeb2017}. Should the data be split in a subject-wise fashion, where all the longitudinal data of each participant is randomly assigned to either the training set or to the test set? Or should the data be split in a record-wise manner, where it is randomly split into training and test sets irrespective of which subjects it belongs to, so that the data from the same subject can belong to both training and test sets?

Recently, Saeb \textit{et al}\cite{saeb2017} demonstrated via simulation studies and the analysis of a publicly available data-set that (for diagnostic applications) splitting the data into training and test sets in a record-wise fashion can lead to massive underestimation of prediction error computed via cross-validation, and that, instead, researchers should adopt subject-wise training/testing data splits. As pointed by the authors, the problem arises because in a record-wise split, the data from each subject can be in both the training and test sets, so that the algorithm is mostly learning about the subject's individual characteristics in the feature data (i.e., it is mostly performing subject identification), instead of learning about the disease characteristics (i.e, performing disease recognition). In other words, the relationship between feature data and disease labels learned by the classifier is confounded by the identity of the subjects (from now on denoted as ``identity confounding"). Now, because the easier task of subject identification replaces the harder task of disease recognition, classifiers trained on data split in a record-wise manner end up achieving overly optimistic prediction accuracy estimates\cite{saeb2017}. (The dangers of record-wise data splitting were independently raised by Sarkar \textit{et al}\cite{sarkar2010,sarkar2013}, who proposed a ``leave-one-individual-out" validation scheme, and by Chaibub Neto \textit{et al}\cite{chaibubneto2017}, where the presence of the identity confounding, denoted as the ``digital fingerprint" in that work, was demonstrated by showing that disease status classifiers were able to achieve high classification accuracies, even when the disease labels were shuffled in a subject-wise fashion.)

In an accompanying discussion article, Little \textit{et al}\cite{little2017} warns about potential issues that might arise from the uncritical adoption of subject-wise cross-validation in diagnostic applications. While there was a consensus that the work of Saeb \textit{et al} demonstrated that identity confounding can be an important issue in practical situations, Little \textit{et al} pointed out situations where subject-wise data splitting violates the assumptions required by cross-validation for the consistent estimation of generalization error. In particular, the adoption of subject-wise splitting with highly heterogeneous data, where it is hard to avoid a mismatch between training and test sets, can make it difficult to train models, and might lead to systematic under-fitting and larger classification errors. Hence, Little \textit{et al} argue that whenever the generalization error estimated with record-wise cross-validation is much lower than with subject-wise cross-validation it is possible that model under-fitting, and not identity confounding, is driving the discrepancy.

Here, we focus on a more manageable problem than the estimation of prediction error via cross-validation, and propose a statistical approach to separate these two situations. We simplify the problem on two fronts. First, we focus on simpler performance metrics than the cross-validation error. (Note that, we don't really need to estimate generalization error in order to show that a classifier performs differently depending on how we split the data. Instead, the comparison of any classification performance metric suffices to show the discrepancy. As a matter of fact, the area under the receiver operating characteristic curve was used in Chaibub Neto \textit{et al}\cite{chaibubneto2017}, to illustrate the discrepancy.) Second, we focus on hypothesis testing, rather than error estimation. Explicitly, we build on the subject-wise label permutation argument used by Chaibub Neto \textit{et al}\cite{chaibubneto2017} to illustrate the presence of identity confounding, and develop two permutation tests. The first test allows us to determine whether the classifier is performing disease recognition, even when identity confounding is present. The second test, builds on the first, and allows us to test for the presence of identity confounding/digital fingerprints per se. By focusing on permutation tests, whose only assumption is the exchangeability of the data under the null hypothesis, we are able to check for identity confounding under more general statistical dependencies and distributional structures of the data, including situations where cross-validation breaks down.

We illustrate the application of the proposed permutation tests with synthetic data examples, as well as, real data from a Parkinson's disease study\cite{bot2016,trister2016}. The synthetic data examples illustrate that identity confounding can arise for many different reasons, including serial dependency in the feature data of each participant (i.e., serially associated records), as well as, due to location and scale differences in the feature distributions across the subjects (even when the longitudinal feature measurements are independent of each other). The real data examples illustrate that a high degree of identity confounding can be present in digital health applications.

The rest of this paper is organized as follows. Section 2 describes the permutation tests. Section 3 illustrates their application to synthetic and real data examples. Finally, Section 4 presents a discussion of our results.

\section{The proposed permutation tests}

Before describing the permutation tests, we first introduce some notation. Throughout the text we let $s = \{1, 2, \ldots, N_s\}$ index the subjects, $r_s = \{1, 2, \ldots, N_{rs} \}$ index the records across subject $s$, and $f = \{1, 2, \ldots, N_f\}$ index the features. $\bfX$ represents the $(N_s \, N_{rs}) \times N_f$ matrix of feature data, with columns indexing the features, and rows indexing the records across all $N_s$ subjects, while $\bfy$ represents the corresponding vector of disease labels, with dimension $(N_s \, N_{rs}) \times 1$. We let $m$ represent an arbitrary classification performance metric, and reserve $p$ to represent the number of permutations employed in our tests. We let $A$ and $B$ represent the events,
\begin{align*}
A &\equiv \mbox{the classifier is performing disease recognition,} \\
B &\equiv \mbox{the classifier is performing subject identification,}
\end{align*}
and adopt standard set theory notation where: $E^c$ represents the complement of event $E$; $\emptyset$ represents the empty set; $\Omega$ represents the universe; and the symbols $\cup$ and $\cap$ represent the union and intersection of events, respectively.

\subsection{Assessing whether the classifier is performing disease recognition}

\begin{figure}[!h]
\begin{center}
\includegraphics[width=\linewidth]{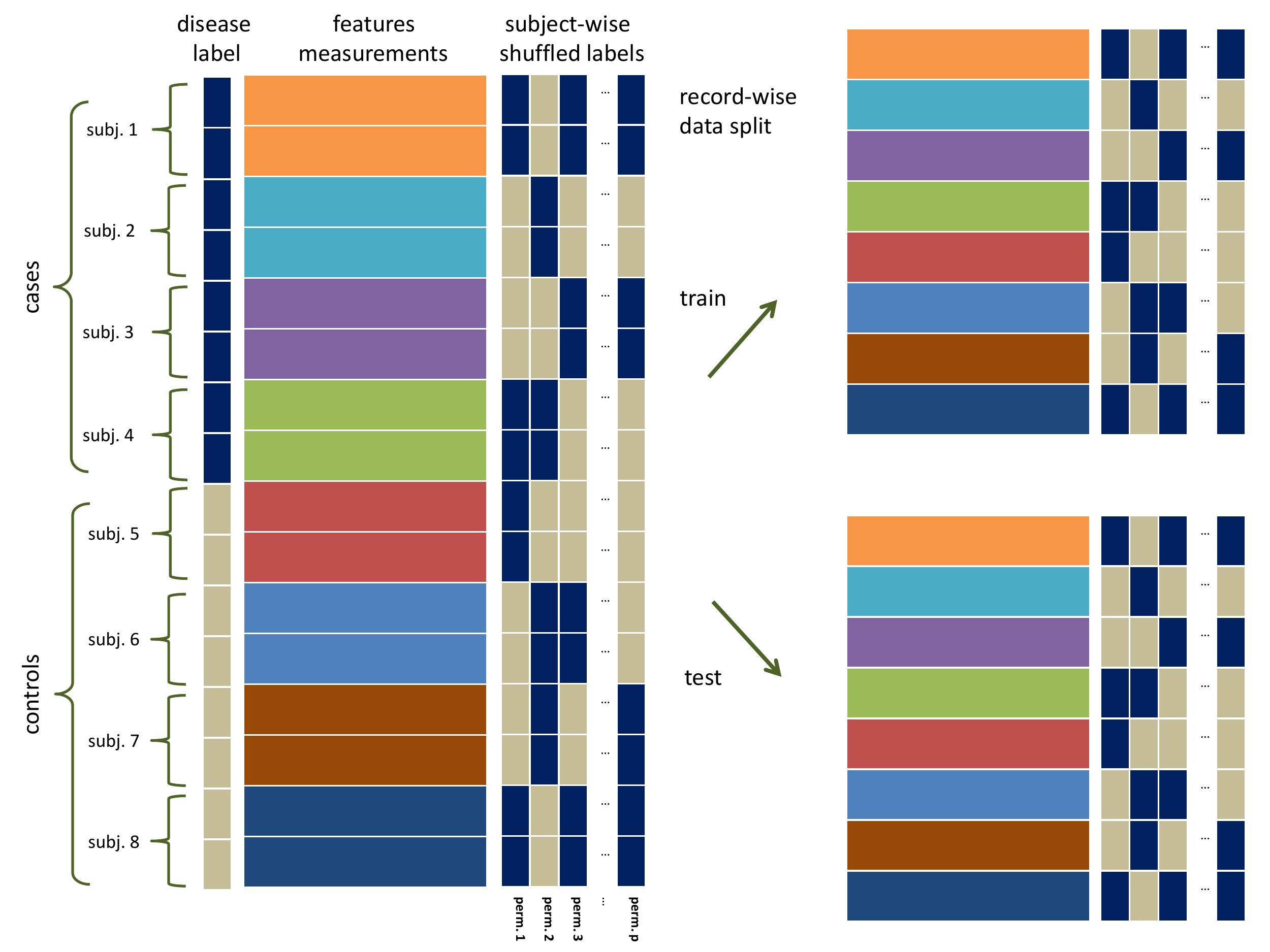}
\caption{Generation of the disease recognition permutation null distribution, using the record-wise data split strategy to create the training and test sets. The schematic shows a few examples of subject-wise shuffled labels, for a data-set with 8 subjects (4 cases and 4 controls), where each subject contributed 2 records. For instance, in the first permutation, the labels of subjects 2 and 3 changed from ``case" to ``control", the labels of subjects 5 and 8 changed from ``control" to ``case", and the labels of subjects 1, 4, 6, and 7 remained the same. (Note that for each subject, the labels are changed across all records.) Adopting the record-wise data split strategy, with one half of the records assigned to the training set, and the other half to the test set, we have that both training and test sets contain 1 sample from each of the subjects.}
\label{fig:recsplitsubjshuffle}
\end{center}
\end{figure}

\begin{figure}[!h]
\begin{center}
\includegraphics[width=\linewidth]{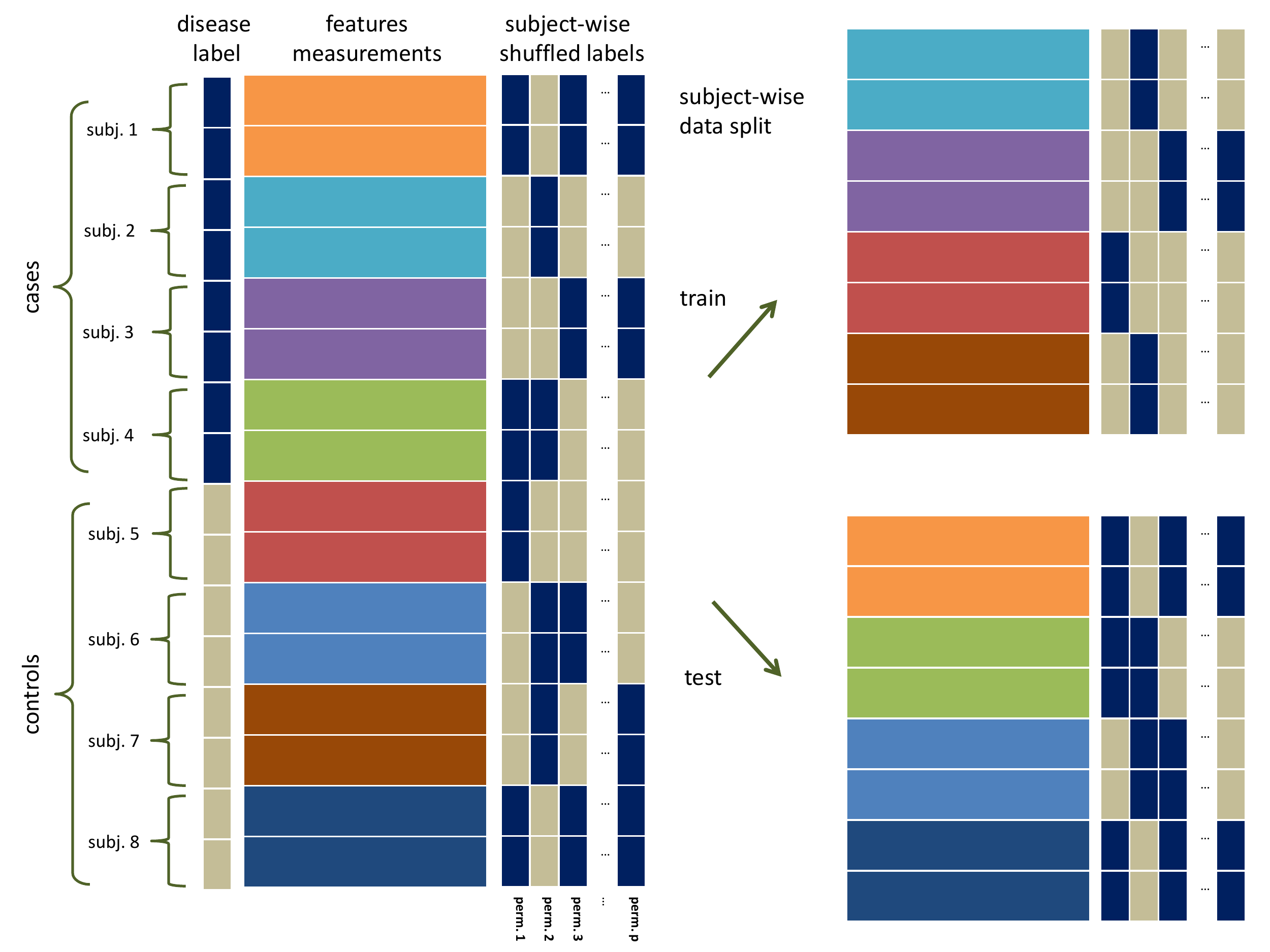}
\caption{Generation of the disease recognition permutation null distribution, using the subject-wise data split strategy to create the training and test sets. The schematic shows a few examples of subject-wise shuffled labels, for a data-set with 8 subjects (4 cases and 4 controls), where each subject contributed 2 records. (Note that for each subject, the labels are changed across all records.) Adopting the subject-wise data split strategy, with half of the subjects assigned to the training set, and the other half to the test set, we have that all the records of each subject are either assigned to the training set, or to the test set.}
\label{fig:subsplitsubjshuffle}
\end{center}
\end{figure}

In order to evaluate whether the classifier is performing disease recognition (even when identity confounding is present) we need to generate a permutation null distribution where the association between the disease labels and the features is destroyed while the association between the features and the subject identities is still preserved. To this end, we generate a permutation distribution by shuffling the disease labels in a subject-wise fashion, as described in Algorithm \ref{alg:permNullDisease}, and illustrated in Figures \ref{fig:recsplitsubjshuffle} and \ref{fig:subsplitsubjshuffle}, and test the hypotheses,
\begin{align*}
H_0^\ast &: \mbox{the classifier is not performing disease recognition,} \\
H_1^\ast &: \mbox{the classifier is performing disease recognition}
\end{align*}

\begin{algorithm}
\caption{Permutation null distribution for disease recognition}\label{alg:permNullDisease}
\begin{algorithmic}[1]
\State \textbf{Input}: Number of permutations, $p$; feature data matrix, $\bfX$; label data vector, $\bfy$; training and test set indexes, $i_{train}$, $i_{test}$
\State Split $\bfX$ into training and test feature sets, $\bfX_{train} \leftarrow \bfX[i_{train},]$, $\bfX_{test} \leftarrow \bfX[i_{test},]$
\For{$i = 1, 2, \ldots, p$}
  \State Generate a subject-wise shuffled version of the label data, $\bfy^{\ast}$
  \State Split $\bfy^{\ast}$ into training and test label sets, $\bfy_{train}^{\ast} \leftarrow \bfy^\ast[i_{train}]$, $\bfy_{test}^{\ast} \leftarrow \bfy^\ast[i_{test}]$
  \State Train a classifier on the $\bfX_{train}$ and $\bfy_{train}^{\ast}$ data
  \State Evaluate the classifier performance on the $\bfX_{test}$ and $\bfy_{test}^{\ast}$ data
  \State Record the value of the performance metric, $m^\ast_i$, on the shuffled data
\EndFor
\State \textbf{Output}: $m^\ast_1$, $m^\ast_2$, \ldots, $m^\ast_p$
\end{algorithmic}
\end{algorithm}

Note that because we are using the performance metric, $m$, as the test statistic of the permutation test (and $m$ is able to capture the contributions of both disease recognition and subject identification to the classification performance), we have that the alternative hypothesis $H_1^\ast$ is equivalent to ``the classifier is performing disease recognition irrespective of whether it is also doing subject identification or not". In other words, $H_1^\ast$ can be more precisely parsed as,
\begin{equation}
H_1^\ast: A \cap (B \cup B^c) = A \cap \Omega = A~,
\end{equation}
while it's complement, the null $H_0^\ast$, can be parsed as,
\begin{equation}
H_0^\ast: (H_1^\ast)^c = A^c \cup (B \cup B^c)^c = A^c \cup (B^c \cap B) = A^c \cup \emptyset = A^c~.
\end{equation}

A permutation p-value for testing $H_0^\ast$ is computed as the proportion of times that the performance metric computed in shuffled label data was better than the performance metric computed with the original (un-shuffled) labels, $m_o$. For instance, for a performance metric such as the area under the receiver operating characteristic curve (AUC), where larger values indicate better classification performance, the permutation p-value can be estimated as
\begin{equation}
\sum_{i=1}^{p} \ind\{m^{\ast}_i \geq m_o\}/p~,
\end{equation}
where $\ind$ represents the indicator function, assuming value 1, when $m^{\ast}_i \geq m_o$, and 0, otherwise. (For performance metrics where smaller values indicate better classification performance, such as misclassification rate, the permutation p-value is given by $\sum_{i=1}^{p} \ind\{m^{\ast}_i \leq m_o\}/p$.)

Figures \ref{fig:recsplitsubjshuffle} and \ref{fig:subsplitsubjshuffle} illustrate the generation of the subject-wise permutation null distribution when the training/test data is split in a record-wise fashion and in subject-wise manner, respectively. Note that when the data is split in a record-wise fashion, and identity confounding is present, the disease recognition null distribution will be centered away from the baseline random guess value (e.g., 0.5 when adopting the AUC) since in the generation of the null distribution the label data is split into training and test sets only after the labels were shuffled (see steps 4 to 7 of Algorithm \ref{alg:permNullDisease}). As a consequence, the presence of identity confounding will show up as an ability to classify the permuted labels with an accuracy that is better than a random guess (since even though the shuffling of labels prevent the algorithm from performing disease recognition, it is still able to perform subject identification). For example, when adopting the AUC metric, and in the presence of strong identity confounding, the disease recognition null will be centered at AUC values closer to 1 than to 0.5. Hence, the disease recognition null distribution not only allows us to test whether the classifier is performing disease recognition, but also informally allows us to infer the presence of identity confounding by simply inspecting how far it is centered from the baseline random guess value. (Observe, nonetheless, that for data split in a subject-wise fashion, the disease recognition null distribution will always be centered around the baseline random guess value, since identity confounding is neutralized by subject-wise data split.) Finally, note that the proposed disease recognition permutation test assumes that the disease labels are exchangeable across subjects (in a subject-wise fashion) under the null $H_0^\ast$.

\subsection{Detecting identity confounding}

As described above, the presence of identity confounding will shift the disease recognition null distribution away from the baseline random guess value. Hence, the median of the disease recognition null distribution,
\begin{equation}
\tilde{m}^\ast \, = \, \mbox{median}(m^{\ast}_1, m^{\ast}_1, \ldots, m^{\ast}_p)~,
\label{eq:obsMedian}
\end{equation}
represents a natural statistic to quantify identity confounding alone, as it measures the contribution of identity confounding to the classifier's predictive ability, after the algorithm's ability to recognize disease has been neutralized by the subject-wise shuffling of disease labels. (Note that while other statistics such as the mean of the disease recognition null distribution could be used to quantify identity confounding, we prefer the more robust median metric since the distribution can be asymmetric.)

\begin{figure}[!h]
\begin{center}
\includegraphics[width=\linewidth]{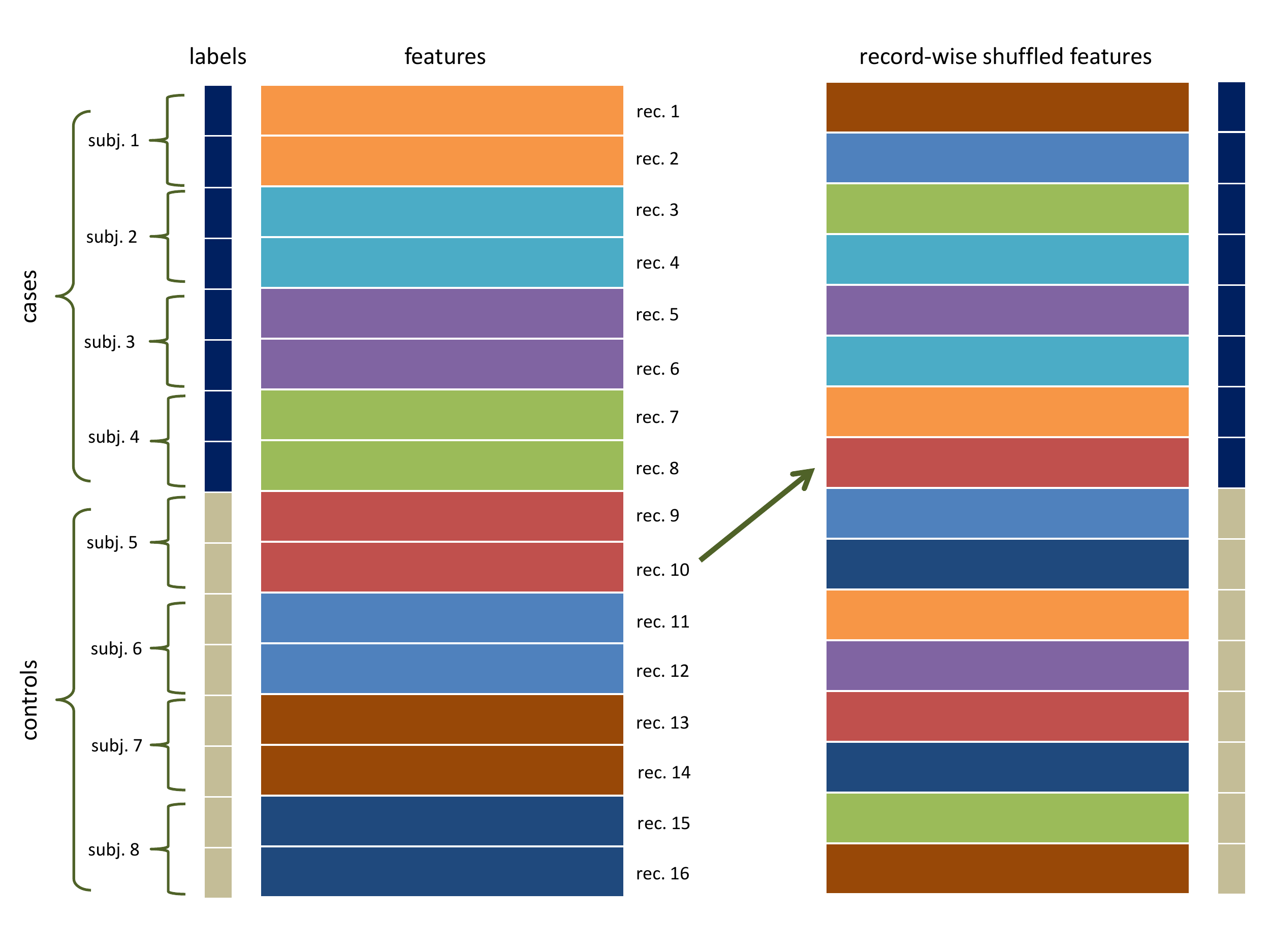}
\caption{Record-wise feature shuffle. The schematic illustrates one instance of record-wise feature shuffling for a data-set containing 8 subjects (4 cases and 4 controls), and 2 records per subject. Record-wise feature shuffling breaks the connection between the features and the subject identities, as well as, between the features and the disease labels. For instance, the green arrow shows that record 10, which was originally associated with subject 5 and a ``control" label in the observed data, is paired with subject 4 and a ``case" label in the shuffled data.}
\label{fig:recordWiseFeatureShuffle}
\end{center}
\end{figure}

In order to test whether the classifier is performing subject identification, we need to generate a permutation null distribution (for the $\tilde{m}^\ast$ statistic) where the association between the subject identities and the features is broken. To this end, we shuffle the feature data in a record-wise fashion, before computing the $\tilde{m}^\ast$ statistic, as described in Algorithm \ref{alg:permNullIndividual}, and illustrated in Figure \ref{fig:recordWiseFeatureShuffle}.
\begin{algorithm}
\caption{Permutation null for detecting identity confounding}\label{alg:permNullIndividual}
\begin{algorithmic}[1]
\State \textbf{Input}: Number of feature permutations, $p$; number of label permutations, $p_l$; feature data matrix, $\bfX$; label data vector, $\bfy$; training and test set indexes, $i_{train}$, $i_{test}$
\For{$i = 1, 2, \ldots, p$}
\State Generate a record-wise shuffled version of the feature data, $\bfX^{\ast}$
\State Split $\bfX^\ast$ into training and test sets, $\bfX^\ast_{train} \leftarrow \bfX^\ast[i_{train},]$, $\bfX^\ast_{test} \leftarrow \bfX^\ast[i_{test},]$
\For{$j = 1, 2, \ldots, p_l$}
  \State Generate a subject-wise shuffled version of the label data, $\bfy^{\ast}$
  \State Split $\bfy^{\ast}$ into training and test label sets, $\bfy_{train}^{\ast} \leftarrow \bfy^\ast[i_{train}]$, $\bfy_{test}^{\ast} \leftarrow \bfy^\ast[i_{test}]$
  \State Train a classifier on the $\bfX^\ast_{train}$ and $\bfy_{train}^{\ast}$ data
  \State Evaluate the classifier performance on the $\bfX^\ast_{test}$ and $\bfy_{test}^{\ast}$ data
  \State Compute the value of the performance metric, $m^{\ast\ast}_j$, on the shuffled data
\EndFor
\State Record the median of the perf. metric distr., $\tilde{m}^{\ast\ast}_i = \mbox{median}(m^{\ast\ast}_1, m^{\ast\ast}_1, \ldots, m^{\ast\ast}_{p_l})$
\EndFor
\State \textbf{Output}: $\tilde{m}^{\ast\ast}_1$, $\tilde{m}^{\ast\ast}_2$, \ldots, $\tilde{m}^{\ast\ast}_p$
\end{algorithmic}
\end{algorithm}

Note that because the record-wise shuffling of the feature data also breaks the association between the features and the disease labels, the identity confounding null distribution will always be centered around the baseline random guess value (0.5 for the AUC metric).

For performance metrics where larger values indicate better classification performance, the permutation p-value for testing the hypotheses,
\begin{align*}
H_0^{\ast\ast} &: \mbox{the classifier is not performing subject identification} \\
H_1^{\ast\ast} &: \mbox{the classifier is performing subject indentification,}
\end{align*}
is estimated as,
\begin{equation}
\sum_{i=1}^{p} \ind\{\tilde{m}^{\ast\ast}_i \geq \tilde{m}^{\ast}\}/p~,
\end{equation}
where $\tilde{m}^{\ast}$ represents the observed value of the test statistic, computed on the original (un-shuffled) feature data, and defined in equation (\ref{eq:obsMedian}), whereas $\tilde{m}^{\ast\ast}_i$ represents the statistic computed on the shuffled feature data, as described in step 12 of Algorithm \ref{alg:permNullIndividual}. (Otherwise, it is computed as $\sum_{i=1}^{p} \ind\{\tilde{m}^{\ast\ast}_i \leq \tilde{m}^{\ast}\}/p$.)

Note that when we shuffle the feature data in a record-wise fashion we, of course, break the connection between the feature data and both the subject identities and the disease labels. We point out, nonetheless, that the null hypothesis $H_0^{\ast\ast}$ is given by $H_0^{\ast\ast}: B^c$ and not by $H_0^{\ast\ast}: B^c \cap A^c$ because, by construction, our test statistic only captures the contribution of the subject identification to the classifier's performance. Observe, as well, that the identity confounding permutation test assumes the exchangeability of the feature data (in a record-wise fashion) under the null $H_0^{\ast\ast}$.

One important drawback of this permutation test is the high amount of computation required to generate the null distribution. We point out, however, that for the particular case where the AUC metric is adopted to evaluate the classifier performance, it is sometimes possible to infer the presence of identity confounding using an analytical approach without having to run the permutation test.

\subsection{An analytical shortcut for detecting identity confounding, based on the AUC metric}

Here, we describe an alternative approach to detect identity confounding. While it is not a proper statistical test, it provides a statistic similar to a p-value (denoted a ``pseudo p-value") which can still be useful in practice. But, before presenting the pseudo p-value statistic, we first describe an additional statistical test that helps understand the pseudo p-value approach.

Consider a third permutation test, where we adopt the AUC as the test statistic, and generate a null distribution by shuffling the disease labels in a record-wise fashion. In this case, because the AUC metric is able to capture the contributions of both disease recognition and subject identification to the classification performance, and the record-wise shuffling of labels breaks the connection between the disease labels and both the feature data and the subject identities, we have that the permutation procedure is now testing the null hypothesis $A^c \cap B^c$ against the alternative $A \cup B$. Hence, the permutation p-value for testing this third set of hypothesis,
\begin{align*}
H_0^{\ast\ast\ast} &: \mbox{the classifier is not performing disease recognition} \\
& \hspace{0.3cm} \mbox{and subject identification,} \\
H_1^{\ast\ast\ast} &: \mbox{the classifier is performing disease recognition,} \\
& \hspace{0.3cm} \mbox{or subject identification.}
\end{align*}
is computed as,
\begin{equation}
\sum_{i=1}^{p} \ind\{\mbox{auc}^{\ast}_i \geq \mbox{auc}_0\}/p~,
\end{equation}
where $\mbox{auc}_0$ represents the AUC value computed from the original data.

For this statistical test, nonetheless, an analytical solution is available. Explicitly, it has been shown\cite{bamber1975} that, when there are no ties in the predicted class probabilities used for the computation of the AUC, the test statistic of the Mann-Whitney U test ($U$) is related to the AUC statistic by, $U = n_n\,n_p (1 - \mbox{AUC})$, where $n_n$ and $n_p$ represent the number of negative and positive labels in the test set (see section 2 of reference\cite{mason2002} for details).

In the presence of ties, the p-value can be computed as the left tail of the asymptotic approximate null,
\begin{equation}
U \, \approx \, \mbox{N}\left( \frac{n_n\,n_p}{2} \; , \; \frac{n_n\,n_p (n + 1)}{12} - \frac{n_n\,n_p}{12 \, n \, (n - 1)} \sum_{j=1}^{\tau} t_j(t_j-1)(t_j+1) \right)~,
\end{equation}
where $n = n_n + n_p$, $\tau$ is the number of groups of ties, and $t_j$ is the number of ties in group $j$\cite{mason2002}.

Alternatively, we can get the p-value as the right tail probability of the corresponding AUC null, $\mbox{AUC} \, \approx \, \mbox{N}(0.5 \, , \, \phi^2)$,
\begin{equation}
\mbox{p-value} \, = \, 1 - \Phi\left( \frac{\mbox{auc}_0 - 0.5}{\phi} \right)~,
\label{eq:null3pval}
\end{equation}
where $\Phi()$ represents the cumulative distribution function of a standard normal distribution, and,
\begin{equation}
\phi^2 \; = \, \frac{n + 1}{12 \, n_n \, n_p} - \frac{1}{12 \, n_n \, n_p \, n \, (n - 1)} \sum_{j=1}^{\tau} t_j(t_j-1)(t_j+1)~.
\end{equation}

Here, we propose to use the pseudo p-value statistic,
\begin{equation}
\mbox{pseudo p-value} \, \equiv \, 1 - \Phi\left( \frac{\tilde{\mbox{auc}}_0^\ast - 0.5}{\phi} \right)~,
\label{eq:pseudopvalue}
\end{equation}
as an alternative measure for identity confounding, where $\tilde{\mbox{auc}}_0^\ast$ corresponds to the median of the disease recognition null distribution computed using the AUC metric.

Note that this procedure does not correspond to a proper statistical test, since it compares the value of one test statistic against the null distribution of a different test statistic, effectively performing an ``apples-to-bananas" comparison (namely, it compares the median of the AUC metric under the disease recognition null distribution against the distribution of the AUC metric under $H_0^{\ast\ast\ast}$). (The correct comparison is the one described in the previous section, where we compare the observed median statistic $\tilde{\mbox{auc}}_0^\ast$ to the null distribution for the same statistic, generated according to Algorithm \ref{alg:permNullIndividual}.)

In any case, the pseudo p-value might be still useful in practice, because the null distribution of the AUC statistic under $H_0^{\ast\ast\ast}$ will always have a larger spread than the null distribution of the median AUC statistic under $H_0^{\ast\ast}$ (since, when we shuffle the feature data in a record-wise fashion, we have that the disease recognition null distribution will be centered at values close to 0.5, so that it's median value will tend to be very close to 0.5). Hence, if the pseudo p-value is already very small, there is no need to perform the computationally expensive permutation test for detecting identity confounding. We point out, however, that a non-significant pseudo p-value does not necessarily mean lack of identity confounding, and that, in this case, it is necessary to compute the permutation p-value since identity confounding can still be present when the pseudo p-value suggests it is not.

\section{Illustrative examples}

We illustrate the application of our permutation tests using synthetic and real data-sets, split into training and test sets according to both the record-wise and subject-wise data split strategies. In all illustrations, we adopt the AUC as the performance metric, and employ the random forest classifier\cite{breimam2001} implemented in the \texttt{randomForest} R package\cite{LiawWiener2002}, using the default tuning parameter specifications.

\subsection{Synthetic data-sets}

When the training and test sets are obtained by record-wise data split, identity confounding can arise due to statistical dependencies in the feature data of each participant (e.g, serial association in longitudinal data), as well as, because of differences in the average or variance of the feature distributions across the distinct subjects (even when the longitudinal feature measurements are independent of each other). Here, we illustrate the application of our permutation tests to synthetic data simulated: (i) in the presence of identity confounding, and absence of disease effect; (ii) in the presence of both identity confounding and disease effect; and (iii) in the absence of both identity confounding and disease effect.

\begin{figure}[!h]
\begin{center}
\includegraphics[width=\linewidth]{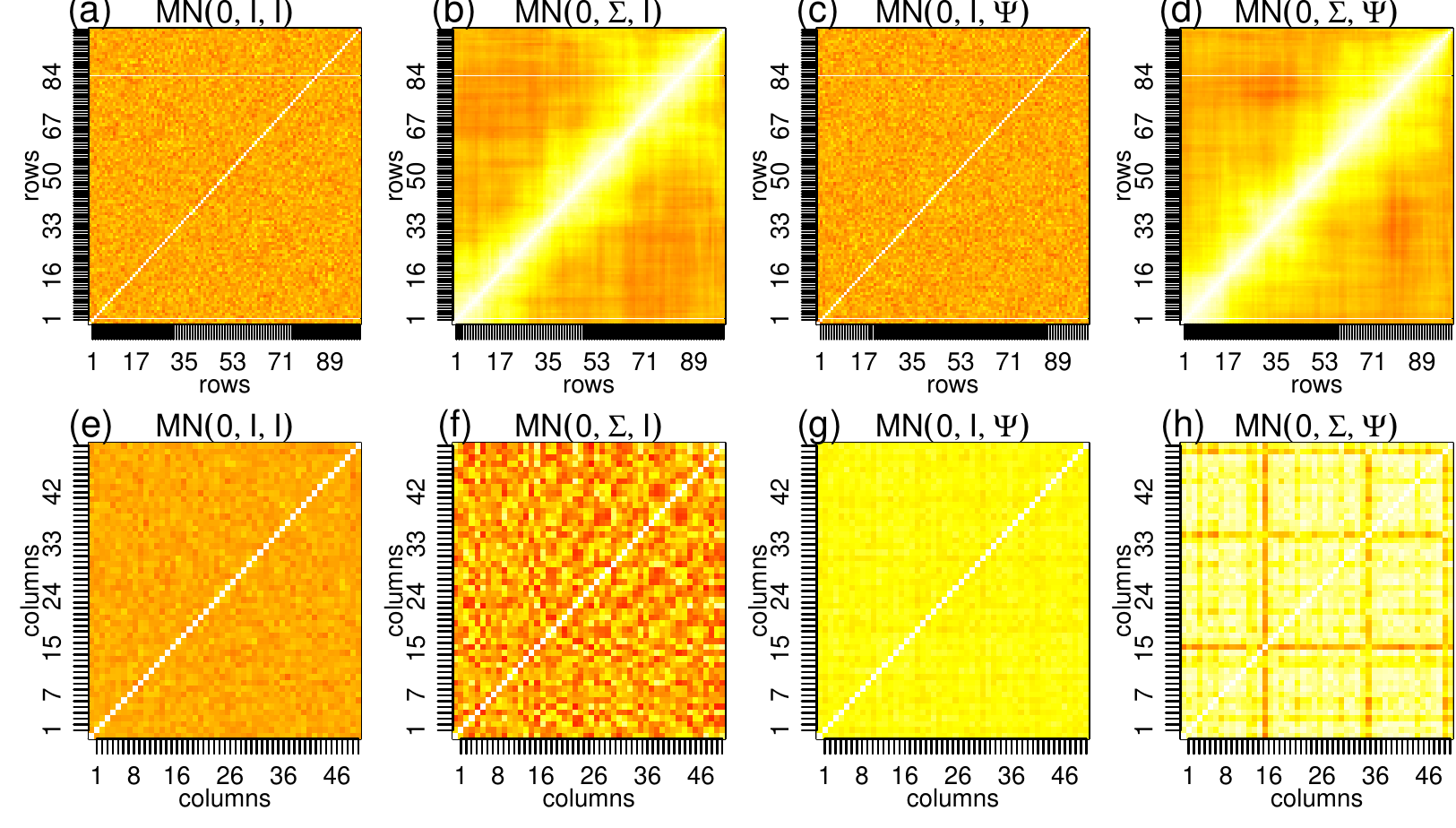}
\caption{Examples of correlation structures generated with the matrix-normal distribution. The top and bottom panels show, respectively, the correlation structure across the rows and columns, for data matrices of dimension $100 \times 50$ sampled from 4 distinct matrix-normal distributions (listed above the heatmaps). Lighter colors represent stronger correlation values. In all examples, we adopted $\rho_r = 0.95$, and $\rho_f = 0.5$ (see Section 3.1). Panels a and e illustrate the case where the data was generated without correlation structures across the rows and columns. Panels b and f depict the case with serial correlation structure across the rows, but no correlation across the columns. Panels c and g, illustrate the case with correlation across the columns, but not across the rows. Finally, panels d and h, depict the case with correlation structure across both rows and columns.}
\label{fig:matrixnormalexamples}
\end{center}
\end{figure}

We simulate feature data using a model similar to the one employed by Saeb \textit{et al}\cite{saeb2017}, except that we generate the data using matrix-normal distributions, in order to model the correlation structure across the records and across the features. (The matrix-normal distribution, over the space of matrices with $r$ rows and $c$ columns, is parametrized as,
\begin{equation}
\bfW \, \sim \, \mbox{MN}_{r \times c}(\bfM \, , \, \bfSigma \, , \, \bfPsi)~,
\end{equation}
where the $r \times c$ matrix $\bfM$ represents the mean of matrix-normal variable $\bfW$; the $r \times r$ matrix $\bfSigma$ models the covariance structure across the rows of $\bfW$; and the $c \times c$ matrix $\bfPsi$ models the covariance structure across the columns of $\bfW$. As an illustration, Supplementary Figure \ref{fig:matrixnormalexamples} shows the correlations across rows and columns for data matrices simulated from 4 distinct matrix-normal distributions.)

In our simulations, the feature data matrix from each subject $s$ was generated independently from all other subjects according to the model,
\begin{equation}
\bfX_s \, = \, \mu_s \, \bfone \,  + \, a \, y_s \, \bfone \, + \, b \, \bfU_s \, + \, c \, \sigma_s \, \bfV_s \, + d \, \bfE_s~,
\label{eq:featuremodel}
\end{equation}
where a, b, c, and d are scalars; $\mu_s$ and $\sigma^2_s$ represent, respectively, subject specific mean and variance values (shared by all features); $\bf1$ corresponds to a matrix with $N_{rs}$ rows and $N_f$ columns filled with 1s; $y_s$ represents the subject's disease label, assuming the value -1 if the subject is a control and 1 if he/she is a case; and the matrices $\bfU_s$, $\bfV_s$, and $\bfE_s$ are sampled from the matrix-normal distributions,
\begin{equation}
\bfU_s \, \sim \, \mbox{MN}_{N_{rs} \times N_f}(\bfzero \, , \, \bfSigma \, , \, \bfI)~,
\end{equation}
\begin{equation}
\bfV_s \, \sim \, \mbox{MN}_{N_{rs} \times N_f}(\bfzero \, , \, \bfI \, , \, \bfI)~,
\end{equation}
\begin{equation}
\bfE_s \, \sim \, \mbox{MN}_{N_{rs} \times N_f}(\bfzero \, , \, \bfI \, , \, \bfPsi)~,
\end{equation}
where we adopt an autoregressive correlation structure for $\bfSigma$ (where each of its elements is given by $\sigma_{ij} = \rho_r^{|i - j|}$, with the fixed scalar $\rho_r$ representing a correlation coefficient), but a simpler correlation structure for $\bfPsi$ (where the diagonal elements are given by 1, and off-diagonal elements are given by the correlation scalar $\rho_f$).

The component $\bfU_s$ induces a correlation structure across the records, while $\bfE_s$ generates correlation across the features. The component $\bfV_s$, on the other hand, only contributes white noise, since by adopting identity matrices for the covariances across the records and the features, we have that sampling from $\mbox{MN}_{N_{rs} \times N_f}(\bfzero , \bfI , \bfI)$ is equivalent to independently sampling each entry of $\bfV_s$ from a standard normal distribution. Also, because $\bfU_s$, $\bfV_s$, and $\bfE_s$ are centered at $\bfzero$, the mean of $\bfX_s$ is given by $\mu_s \, \bfone  + a \, y_s \, \bfone$.

\begin{figure}[!h]
\begin{center}
\includegraphics[width=\linewidth]{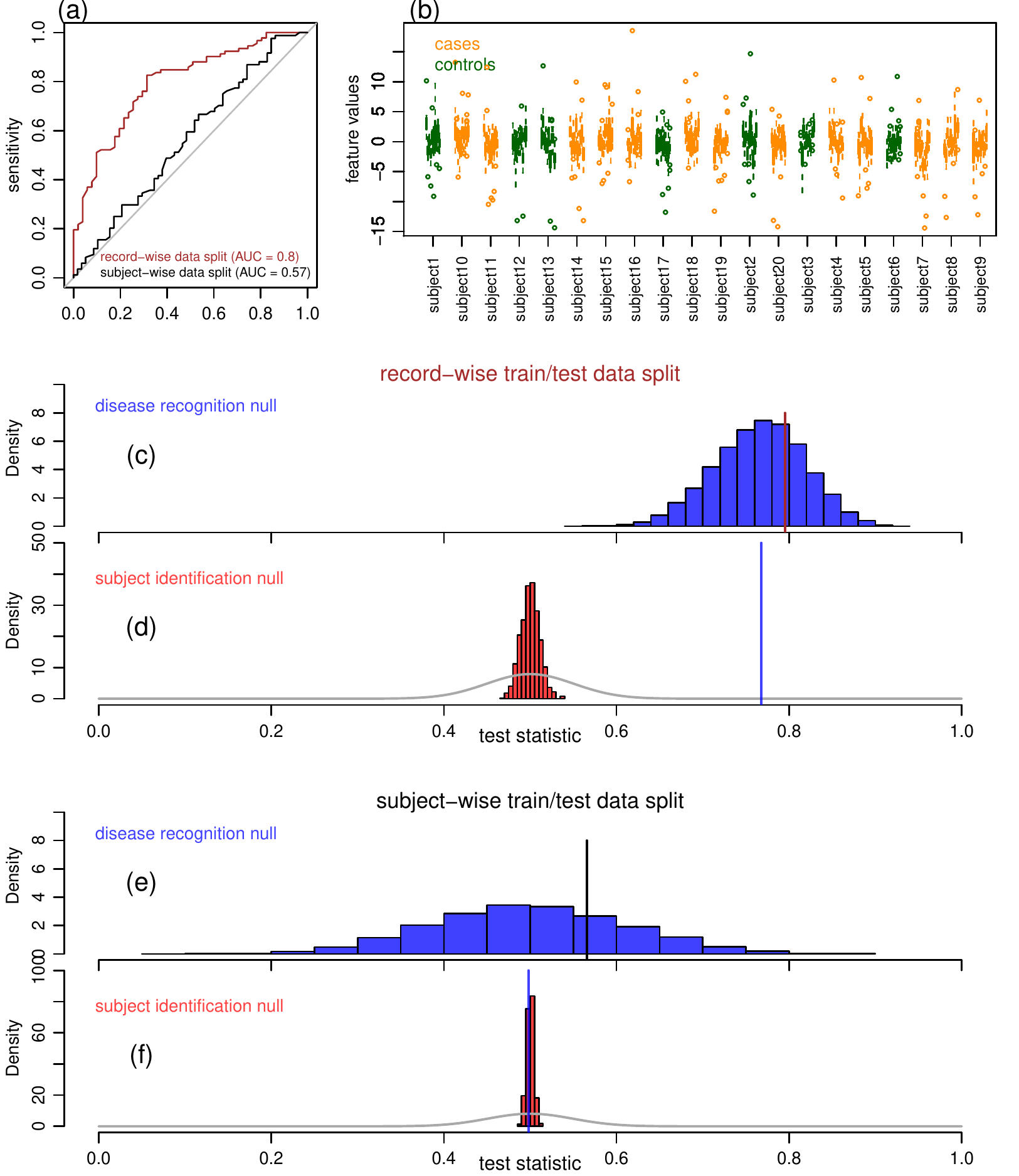}
\caption{Synthetic data example 1, where the feature data was simulated from model $\bfX_s = 2 \, \bfU_s + \bfV_s + 0.5 \, \bfE_s$, under the null hypothesis for disease recognition, $H_0^{\ast}$ and the alternative hypothesis for identity confounding, $H_1^{\ast\ast}$. Panel a shows the ROC curves and AUC values for data split in a record-wise (brown) and subject-wise (black) fashion. Panel b shows boxplots of the feature data across all subjects, with cases and controls shown in orange and green, respectively. Panels c and d show, respectively, the disease recognition and the identity confounding permutation null distributions (blue and red histograms) for the record-wise data split. The brown line in panel c corresponds to the observed AUC value, while the blue line and grey curve in panel d show, respectively, the median of the blue histogram in panel c and the density of the normal distribution used for the computation of the pseudo p-value. Panels e and f show the analogous objects for the subject-wise data split.}
\label{fig:example1}
\end{center}
\end{figure}

\begin{figure}[!h]
\begin{center}
\includegraphics[width=\linewidth]{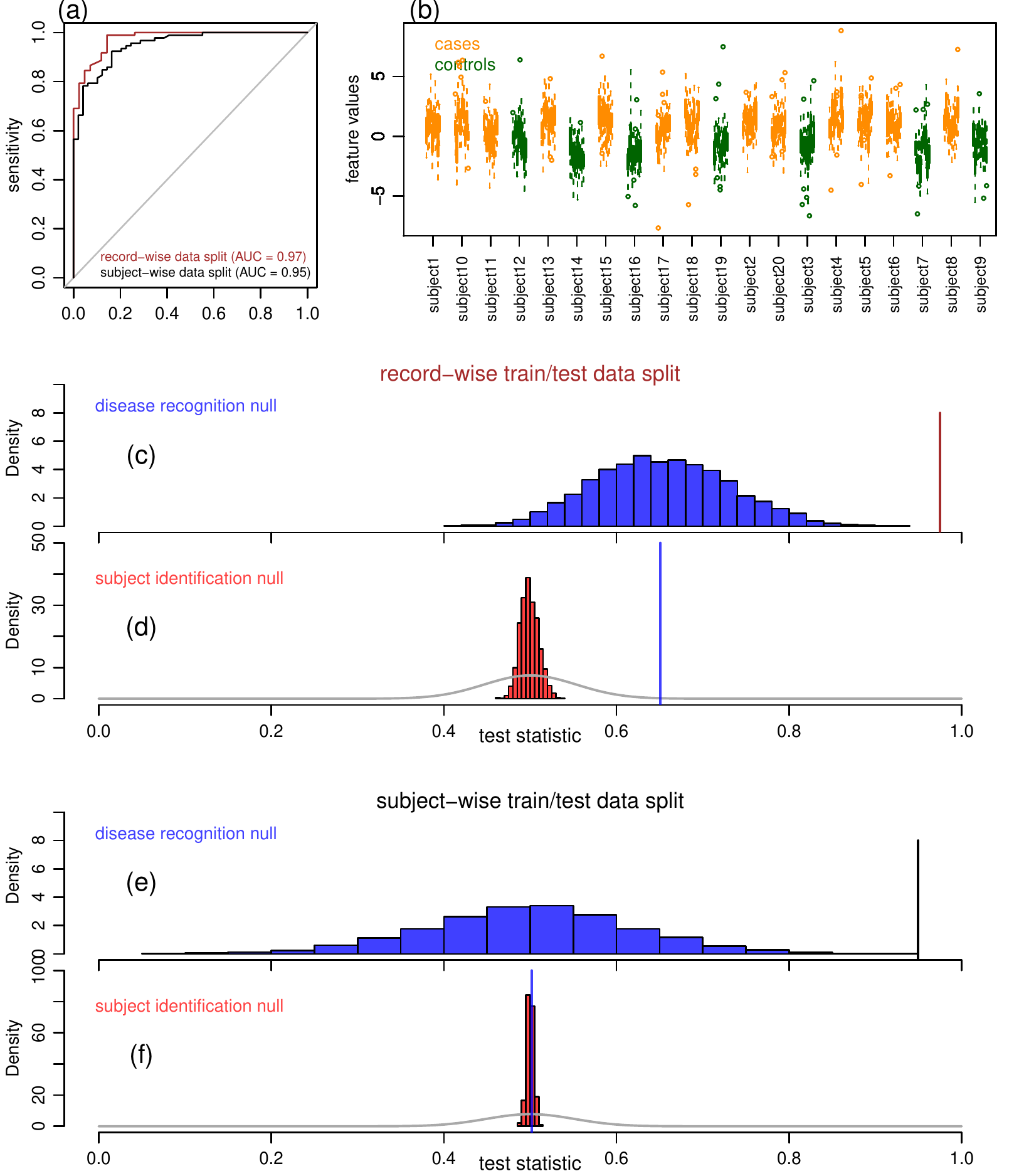}
\caption{Synthetic data example 2, where the feature data is simulated from model $\bfX_s = y_s \, \bfone + 2 \, \bfU_s + \bfV_s + 0.5 \, \bfE_s$, under the alternative hypothesis for disease recognition, $H_1^{\ast}$, and the alternative hypothesis for identity confounding, $H_1^{\ast\ast}$. Panel a shows the ROC curves and AUC values for data split in a record-wise (brown) and subject-wise (black) fashion. Panel b shows boxplots of the feature data across all subjects, with cases and controls shown in orange and green, respectively. Panels c and d show, respectively, the disease recognition and the identity confounding permutation null distributions (blue and red histograms) for the record-wise data split. The brown line in panel c corresponds to the observed AUC value, while the blue line and grey curve in panel d show, respectively, the median of the blue histogram in panel c and the density of the normal distribution used for the computation of the pseudo p-value. Panels e and f show the analogous objects for the subject-wise data split.}
\label{fig:example2}
\end{center}
\end{figure}

\begin{figure}[!h]
\begin{center}
\includegraphics[width=\linewidth]{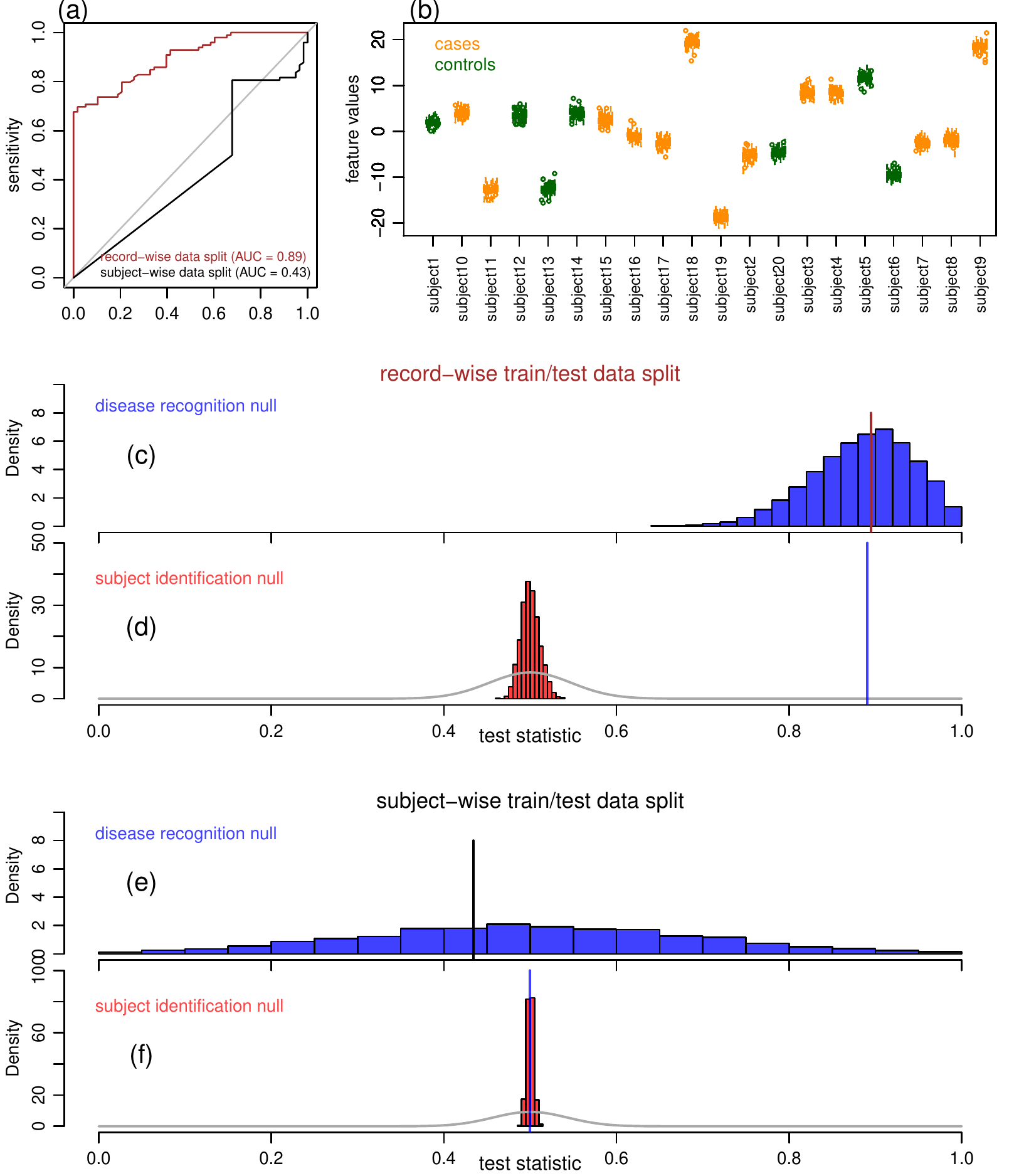}
\caption{Synthetic data example 3, where the feature data is simulated from model $\bfX_s = \mu_s \, \bfone + \bfV_s$, under the null hypothesis for disease recognition, $H_0^{\ast}$, and the alternative hypothesis for identity confounding, $H_1^{\ast\ast}$. Panel a shows the ROC curves and AUC values for data split in a record-wise (brown) and subject-wise (black) fashion. Panel b shows boxplots of the feature data across all subjects, with cases and controls shown in orange and green, respectively. Panels c and d show, respectively, the disease recognition and the identity confounding permutation null distributions (blue and red histograms) for the record-wise data split. The brown line in panel c corresponds to the observed AUC value, while the blue line and grey curve in panel d show, respectively, the median of the blue histogram in panel c and the density of the normal distribution used for the computation of the pseudo p-value. Panels e and f show the analogous objects for the subject-wise data split.}
\label{fig:example3}
\end{center}
\end{figure}

\begin{figure}[!h]
\begin{center}
\includegraphics[width=\linewidth]{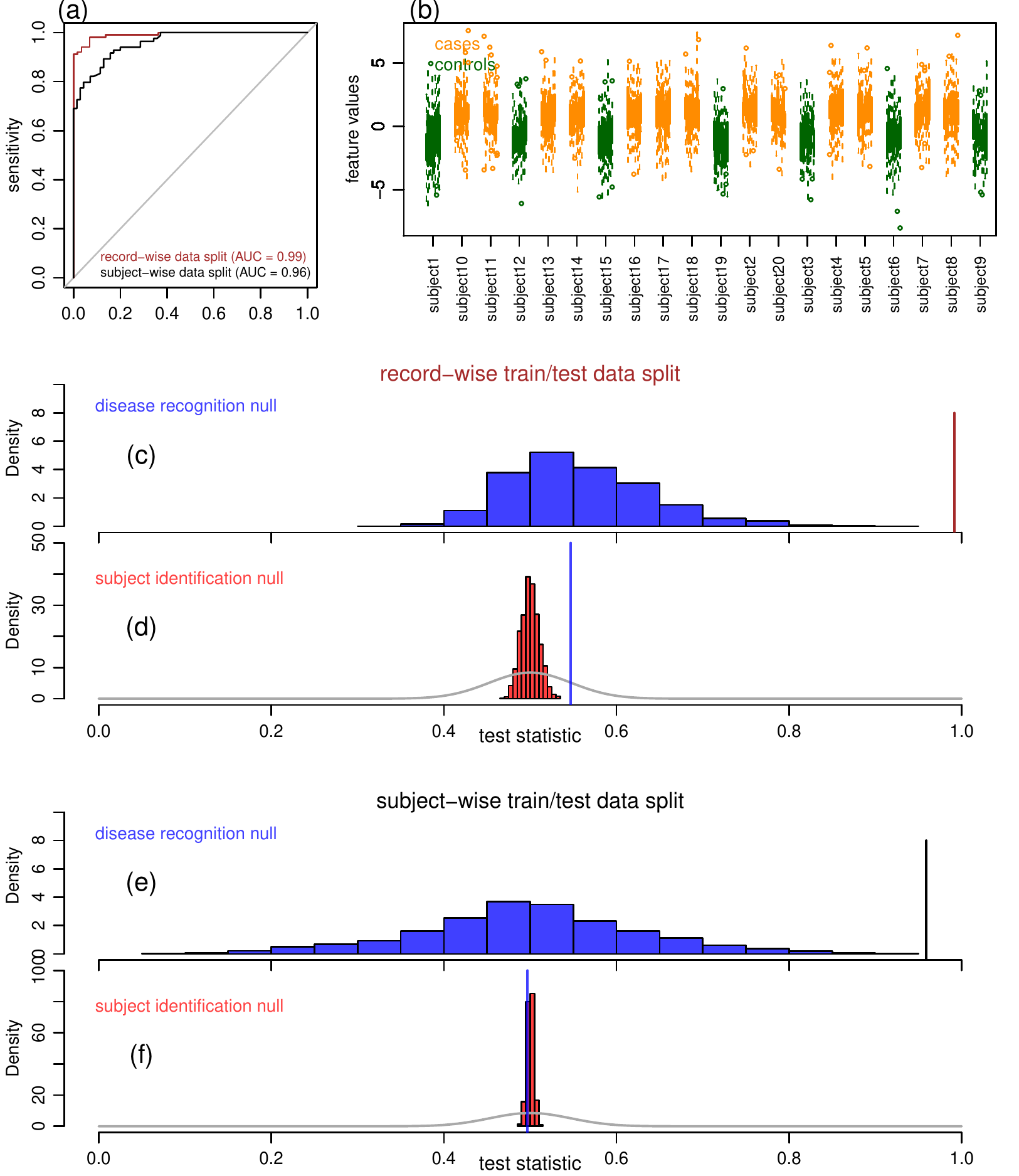}
\caption{Synthetic data example 4, where the feature data is simulated from model $\bfX_s = y_s \, \bfone + \bfV_s$, under the alternative hypothesis for disease recognition, $H_1^{\ast}$, and the alternative hypothesis for identity confounding, $H_1^{\ast\ast}$. Panel a shows the ROC curves and AUC values for data split in a record-wise (brown) and subject-wise (black) fashion. Panel b shows boxplots of the feature data across all subjects, with cases and controls shown in orange and green, respectively. Panels c and d show, respectively, the disease recognition and the identity confounding permutation null distributions (blue and red histograms) for the record-wise data split. The brown line in panel c corresponds to the observed AUC value, while the blue line and grey curve in panel d show, respectively, the median of the blue histogram in panel c and the density of the normal distribution used for the computation of the pseudo p-value. Panels e and f show the analogous objects for the subject-wise data split.}
\label{fig:example4}
\end{center}
\end{figure}

\begin{figure}[!h]
\begin{center}
\includegraphics[width=\linewidth]{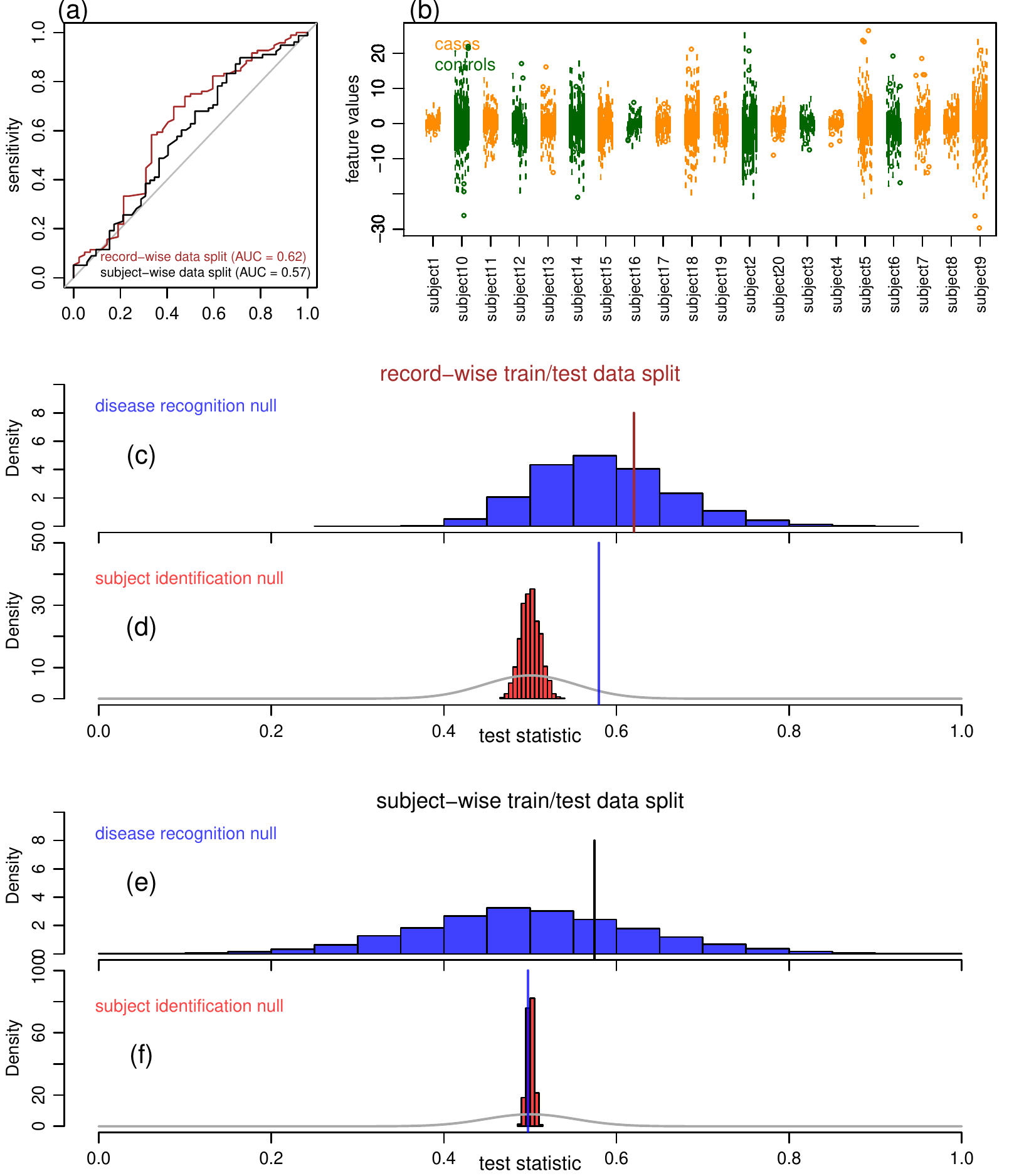}
\caption{Synthetic data example 5, where the feature data is simulated from model $\bfX_s = \sigma_s \, \bfV_s$, under the null hypotheses for disease recognition, $H_0^{\ast}$, and identity confounding, and the alternative hypothesis for identity confounding, $H_1^{\ast\ast}$. Panel a shows the ROC curves and AUC values for data split in a record-wise (brown) and subject-wise (black) fashion. Panel b shows boxplots of the feature data across all subjects, with cases and controls shown in orange and green, respectively. Panels c and d show, respectively, the disease recognition and the identity confounding permutation null distributions (blue and red histograms) for the record-wise data split. The brown line in panel c corresponds to the observed AUC value, while the blue line and grey curve in panel d show, respectively, the median of the blue histogram in panel c and the density of the normal distribution used for the computation of the pseudo p-value. Panels e and f show the analogous objects for the subject-wise data split.}
\label{fig:example5}
\end{center}
\end{figure}

\begin{figure}[!h]
\begin{center}
\includegraphics[width=\linewidth]{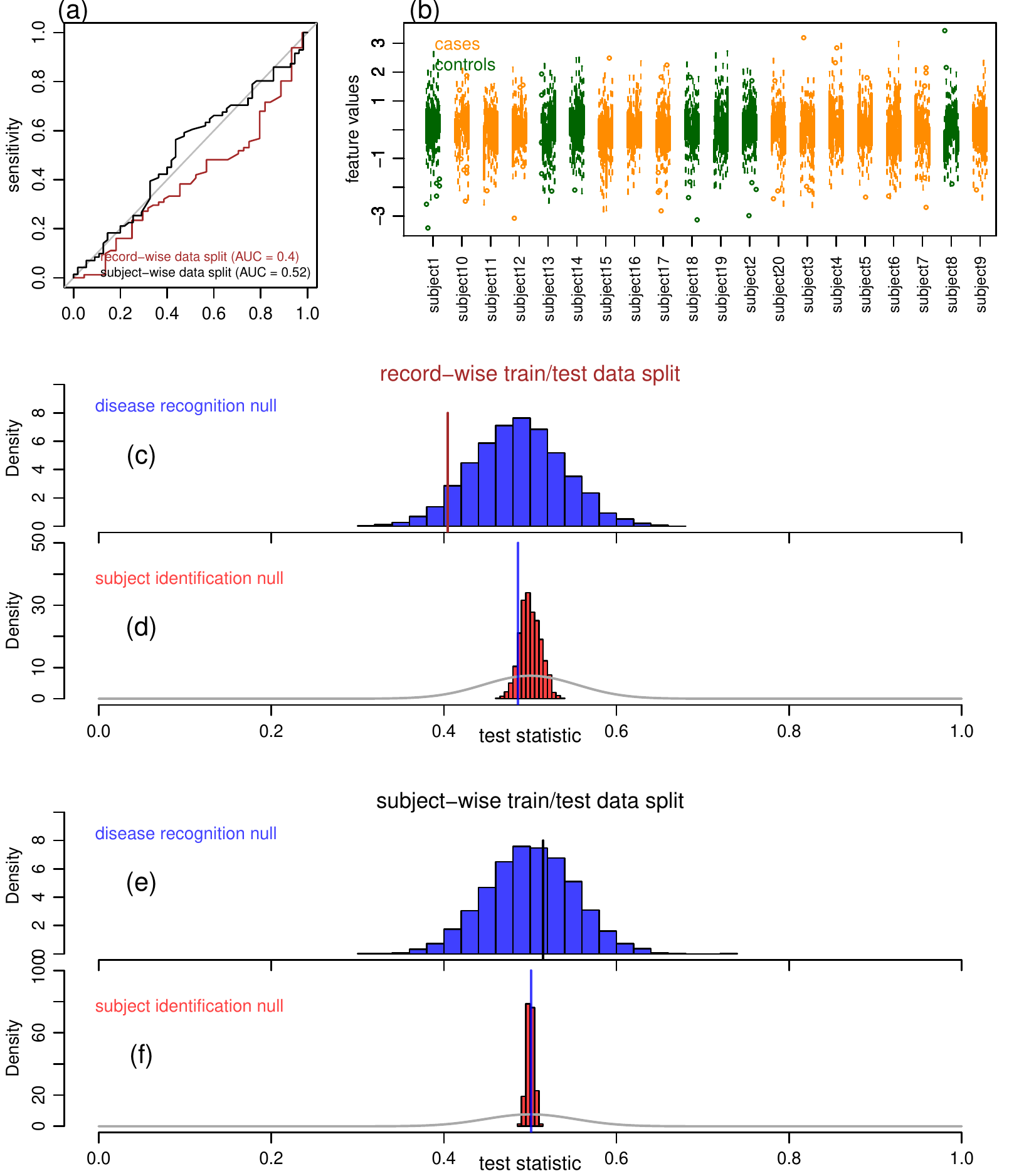}
\caption{Synthetic data example 6, where the feature data is simulated from model $\bfX_s = \bfV_s$, under the null hypotheses for disease recognition, $H_0^{\ast}$, and identity confounding, $H_0^{\ast\ast}$. Panel a shows the ROC curves and AUC values for data split in a record-wise (brown) and subject-wise (black) fashion. Panel b shows boxplots of the feature data across all subjects, with cases and controls shown in orange and green, respectively. Panels c and d show, respectively, the disease recognition and the identity confounding permutation null distributions (blue and red histograms) for the record-wise data split. The brown line in panel c corresponds to the observed AUC value, while the blue line and grey curve in panel d show, respectively, the median of the blue histogram in panel c and the density of the normal distribution used for the computation of the pseudo p-value. Panels e and f show the analogous objects for the subject-wise data split.}
\label{fig:example6}
\end{center}
\end{figure}

By varying the scalar values $\mu_s$, $\sigma_s$, $a$, $b$, $c$, and $d$ in equation (\ref{eq:featuremodel}), we can simulate data under the null and alternative hypothesis for disease recognition and identity confounding. Here, we illustrate the application of our permutation tests in 6 distinct scenarios presented in Figures \ref{fig:example1}, \ref{fig:example2}, \ref{fig:example3}, \ref{fig:example4}, \ref{fig:example5}, and \ref{fig:example6}. Table \ref{tab:syntheticexamples} provides a brief description of the mechanisms giving rise to identity confounding, the hypotheses under which the data was simulated, and the models used to generate the feature data. Our goal is to illustrate how identity confounding can arise for many different reasons, and how the proposed permutation tests are able to detect it.

\begin{table}[!h]
{\scriptsize
\begin{tabular}{clll}
\hline
Example & Identity confounding  & Hypotheses ($H_0$ or $H_1$)  & Model used to simulate \\
(Figure) & generation mechanism & consistent with the data & the feature data, $\bfX_s$ \\
\hline \hline
Example 1 & serial dependency & disease recog. null ($H_0^\ast$), & $2 \, \bfU_s + \bfV_s + 0.5 \, \bfE_s$ \\
(Fig. \ref{fig:example1}) & across records ($\bfU_s$) & identity conf. alter. ($H_1^{\ast\ast}$) & \\
\hline
Example 2 & serial dependency & disease recog. alter. ($H_1^\ast$),  & $y_s \, \bfone + 2 \, \bfU_s + \bfV_s + 0.5 \bfE_s$ \\
(Fig. \ref{fig:example2}) & across records ($\bfU_s$), & identity conf. alter. ($H_1^{\ast\ast}$) & \\
& and location shifts & & \\
& due to disease label ($y_s$) & & \\
\hline
Example 3 & location shifts due & disease recog. null ($H_0^\ast$),  & $\mu_s \, \bfone + \bfV_s$ \\
(Fig. \ref{fig:example3}) & to subject specific & identity conf. alter. ($H_1^{\ast\ast}$) & \\
& mean ($\mu_s$) & & \\
\hline
Example 4 & location shifts due & disease recog. alter. ($H_1^\ast$),  & $y_s \, \bfone + \bfV_s$ \\
(Fig. \ref{fig:example4}) & to disease label ($y_s$) & identity conf. alter. ($H_1^{\ast\ast}$) & \\
\hline
Example 5 & scale shifts due & disease recog. null ($H_0^\ast$),  & $\sigma_s \, \bfV_s$ \\
(Fig. \ref{fig:example5}) & to subject specific & identity conf. alter. ($H_1^{\ast\ast}$) & \\
& variances ($\sigma_s^2$) & & \\
\hline
Example 6 & no identity confounding & disease recog. null ($H_0^\ast$),  & $\bfV_s$ \\
(Fig. \ref{fig:example6}) & generation mechanism & identity conf. null ($H_0^{\ast\ast}$) & \\
\hline
\end{tabular}}
\caption{Summary of the synthetic data examples. The first column lists the synthetic data examples (and the respective figures where they are presented). The second column describes the mechanisms giving rise to identity confounding in each example. The third describes which disease recognition hypotheses ($H_0^\ast$ vs $H_1^\ast$) and identity confounding hypothesis ($H_0^{\ast\ast}$ vs $H_1^{\ast\ast}$) are consistent with each simulated example. The last column describes the respective models used of simulate the feature data. Further details are provided in the main text.}
\label{tab:syntheticexamples}
\end{table}

In all examples, we simulated data for 10 correlated features ($\rho_f = 0.5$), across 20 subjects (13 cases and 7 controls), with the number of records varying from 10 to 20 records per subject, and adopting $\rho_r = 0.95$. We employed 10,000 permutations in the generation of the disease recognition null distribution. The generation of the identity confounding null distribution, on the other hand, was based on only 1,000 record-wise feature permutations, with 300 subject-wise label permutations per record-wise feature permutation due to computational constraints. (Supplementary Figure S2, provides a justification for our choice of 300 subject-wise label shufflings. It shows that the sampling variability associated with the adoption of 1,000 record-wise feature permutations overwhelms the small differences in the shape of the identity confounding null achieved by increasing the number of subject-wise label shufflings beyond 300 permutations.)

In our first synthetic data example, we simulated feature data from the model,
\begin{equation}
\bfX_s \, = \, 2 \, \bfU_s \, + \, \bfV_s \, + 0.5 \, \bfE_s~,
\label{eq:example1}
\end{equation}
where identity confounding arises due to the serial correlation across the records induced by the $\bfU_s$ component. Since, the model does not include a disease effect ($a$ was set to 0), we have that the data is consistent with the null hypothesis of the disease recognition permutation test, $H_0^\ast$, and with the alternative hypothesis of the identity confounding permutation test, $H_1^{\ast\ast}$. Figure \ref{fig:example1}a shows the ROC curves for data split in a record-wise (brown) and subject-wise (black) fashion. Note that the subject-wise split leads to a much smaller AUC value (0.57) than the record-wise data split (AUC $= 0.8$). Figure \ref{fig:example1}b shows boxplots of the feature data across all 20 subjects, with cases and controls shown in orange and green, respectively. In this case, we don't see mean differences between cases and controls, since we simulated featured data centered at 0 (both $\mu_s$ and $a$ are set to 0 in the model shown in equation \ref{eq:example1}). Figure \ref{fig:example1}c shows the disease recognition null distribution (blue), and the observed AUC value, $auc_o$, (brown line) for the record-wise data split. The null distribution is centered away from 0.5 (median $= 0.77$), indicating the presence of identity confounding. The disease recognition permutation p-value ($= 0.2984$, given by the tail probability of the blue distribution to the right of the brown line), nonetheless, indicates that the classifier is still unable to perform disease recognition, even though the observed AUC value (0.8) is high. This observation suggests that the high classification performance is entirely due to the classifier's ability to identify subjects. Figure \ref{fig:example1}d, shows the identity confounding null distribution (red), and the observed value of the $\tilde{auc}^\ast$ statistic (blue line), which corresponds to the median of the disease recognition null distribution shown in panel c. The identity confounding permutation p-value (computed as the proportion of times that the red distribution was larger than the blue line) is 0, and confirms the presence of identity confounding when the data is  split in a record-wise fashion. The grey density represents the analytical approximation for the $H_0^{\ast\ast\ast}$ null, used in the computation of the the pseudo p-value for detecting identity confounding (given by the area under the grey density to the right of the blue line - which, in this example, is $6 \times 10^{-8}$). Figure \ref{fig:example1}e shows the disease recognition null distribution for the subject-wise data split (blue) and the observed AUC value (black line). Because the subject-wise split (by construction) eliminates the classifier's ability to perform subject identification, this null distribution will always be centered around 0.5. In this example, again, the disease recognition permutation p-value (0.2781) indicates that the random forest classifier is is not performing disease recognition. Figure \ref{fig:example1}f shows the identity confounding null distribution (red) for the subject-wise data split, and further confirms that this data split avoids identity confounding issues (permutation p-value $= 0.719$, pseudo p-value $= 0.516$).

In our second synthetic data example, we simulate data from the model,
\begin{equation}
\bfX_s \, = \, y_s \, \bfone\, + \, 2 \, \bfU_s \, + \, \bfV_s \, + 0.5 \, \bfE_s~,
\label{eq:example2}
\end{equation}
which basically adds a disease effect to the model used in the first example. Hence, the data is now simulated under the alternative hypothesis for disease recognition, $H_1^\ast$, and identity confounding, $H_1^{\ast\ast}$. Figure \ref{fig:example2}a shows much higher AUC values for both record-wise (0.97) and subject-wise (0.95) data splits. Figure \ref{fig:example2}b shows clear mean differences between cases and controls (as one would expect, since we now include a disease effect). Figure \ref{fig:example2}c shows that, even though identity confounding is playing a role in the random forest's classification performance, when the data is split in a record-wise manner (note the shift away from 0.5 in the blue distribution), the algorithm is, nonetheless, still doing disease recognition (permutation p-value $= 0$). Figure \ref{fig:example2}e, shows that the random forest is performing disease recognition with the subject-wise data split, as well (permutation p-value $= 0$). Figure \ref{fig:example2}d confirms the presence of identity confounding (permutation p-value $= 0$, pseudo p-value $= 0.0024$) with record-wise data split, while Figure \ref{fig:example2}f confirms it's absence (permutation p-value $= 0.318$, pseudo p-value $= 0.487$) with subject-wise split.

In our third example, we illustrate how differences in the mean feature values across the subjects can lead to identity confounding, even when the longitudinal data of each individual has no serial association structure. To this end, we simulate data from the model,
\begin{equation}
\bfX_s \, = \, \mu_s \, \bfone \, + \, \bfV_s~,
\label{eq:example3}
\end{equation}
where the feature data of each subject is independent and identically distributed (i.i.d.) according to $\mbox{N}(\mu_s, 1)$. Note that in this case the data is simulated under the null for disease recognition, $H_0^\ast$, and under the alternative for identity confounding, $H_1^{\ast\ast}$. (Observe, as well, that this model corresponds to the i.i.d. mixture model for data clustered by subject, discussed in reference\cite{little2017}.) Figure \ref{fig:example3}b shows accentuated mean differences in feature values across the subjects, and Figures \ref{fig:example3}c and d illustrate how, even though, the record-wise data split leads to a high AUC value (0.89), the random forest classifier is performing subject identification only (disease recognition permutation p-value $= 0.474$, identity confounding permutation p-value $= 0$).

Our fourth example, illustrates how the disease effect can itself induce a certain amount of identity confounding, when the data is split in a record-wise manner. To this end, we simulated data from the model,
\begin{equation}
\bfX_s \, = \, y_s \, \bfone \, + \, \bfV_s~,
\label{eq:example4}
\end{equation}
where the feature data from case and control subjects is i.i.d. $\mbox{N}(1, 1)$ and $\mbox{N}(-1, 1)$, respectively. Because the disease label contributes to the mean of the feature values, and identity confounding can arise because of mean differences across the subjects (as illustrated in the previous example), we have that, the data in model (\ref{eq:example4}) is generated under the alternative for disease recognition, $H_1^{\ast}$, as well as, for identity confounding, $H_1^{\ast\ast}$. (Observe that, at least for data simulated according to model (\ref{eq:featuremodel}), it is not possible to generate data under the alternative for disease recognition and, concomitantly, under the null for identity confounding.) Figure \ref{fig:example4}b shows clear mean differences between cases and controls, while Figure \ref{fig:example4}c shows that the classifier is performing disease recognition with the record-wise data split (permutation p-value $= 0$). But, more interestingly, it also illustrates that the disease recognition null is slightly shifted away from 0.5 (median $= 0.55$), showing that the difference in means is enough to generate identity confounding for record-wise data splits. Furthermore, Figure \ref{fig:example4}d shows that this small shift is statistically significant (identity confounding permutation p-value $= 0$). Observe, as well, that this example illustrates the case where the pseudo p-value ($= 0.162$) fails to detect identity confounding, while the permutation p-value does not.

In our fifth example, we illustrate how differences in the variance of the feature values across the subjects can lead to identity confounding, even when the longitudinal data of each individual has no serial association structure. To this end, we simulate data from the model,
\begin{equation}
\bfX_s \, = \, \sigma_s \, \bfV_s~,
\label{eq:example5}
\end{equation}
where the feature data of each subject is i.i.d. according to $\mbox{N}(0, \sigma^2_s)$, and the $\sigma^2_s$ values were sampled from a uniform distribution in the interval $(1, 10)$. Note that in this case the data is again simulated under the null for disease recognition, $H_0^\ast$, but under the alternative for identity confounding, $H_1^{\ast\ast}$. Figure \ref{fig:example5}b shows clear spread differences between the subjects. Note that although Figure \ref{fig:example5}c shows that the classifier is not performing disease recognition with the record-wise data split (permutation p-value $= 0.3088$), it illustrates that the difference in variance is enough to generate identity confounding for record-wise data splits. Furthermore, Figure \ref{fig:example5}d shows that this small shift is statistically significant (identity confounding permutation p-value $= 0$). Note, as well, that this example illustrates the case where the pseudo p-value ($= 0.069$) is marginally significant, while the permutation p-value clearly indicates the presence of identity confounding.

In our sixth example, we consider the null model,
\begin{equation}
\bfX_s \, = \, \bfV_s~,
\label{eq:example6}
\end{equation}
where all the feature data within and across all participants is i.i.d. $\mbox{N}(0, 1)$, so that the data is simulated under the null hypothesis for disease recognition, $H_0^{\ast}$, and identity confounding, $H_0^{\ast\ast}$. As expected, Figure \ref{fig:example6} shows that the random forest classifier is not performing disease recognition or subject identification, no matter how the data is split.

Finally, as a sanity check, we performed a simulation study using 500 distinct data-sets simulated under the null for both disease recognition and identity confounding. Supplementary Figure S3 reports the results and describes the simulation experiment in more detail. As expected, the distribution of the p-values is approximately uniform for the permutation tests, indicating that the type I error rates are being controlled at the nominal significance levels. The pseudo p-value distributions, on the other hand, are bell shaped and centered around 0.5, indicating that the pseudo p-value approach is conservative at small nominal significance levels (that is, the pseudo p-values tend to be larger than they should be), illustrating, once again, that a small pseudo p-value implies the presence of identity confounding, whereas non-significant pseudo p-values do not imply the absence of digital fingerprints.

\subsection{Real data illustrations}

We illustrate the application of our proposed tests to the diagnosis of Parkinson's disease (PD), using iPhone sensor data collected during the first six months of the mPower study\cite{trister2016,bot2016}. We focus our analyses on data collected from the voice and tapping activity tasks (details about the activity tasks are provided in reference\cite{bot2016}).

Our analyses were based on a matched sample of 11 PD patients and 11 control participants, who provided at least 100 records over the 6 month period. All case and control participants were males, and were first matched by age (using exact matching), and then by education level (using nearest neighbor matching\cite{ho2011}) whenever there were multiple cases with the same age of a control, or vice-versa. (In the event that ties persisted after the application of this second matching criterion, we randomly selected one participant to perform the matching.) We trained separate random forest classifiers for the voice and tapping data, based on 13 voice features and 41 tapping features proposed in the literature\cite{tsanas2011,arora2015}.

Due to the small number of subjects, in order to maintain balance in the age distributions of cases and controls in the training and test sets, we performed an age balanced version of the subject-wise data split (where whenever the training set contained a case of a given age, then the test set contained the matched control of the same age, and vice versa). For the record-wise data split, on the other hand, we simply randomly selected the records. (Note that because each subject has a large number of records (at least one hundred), the distribution of the ages tends to be well balanced between cases and controls in both the training and test sets.)

\begin{figure}[!h]
\begin{center}
\includegraphics[width=\linewidth]{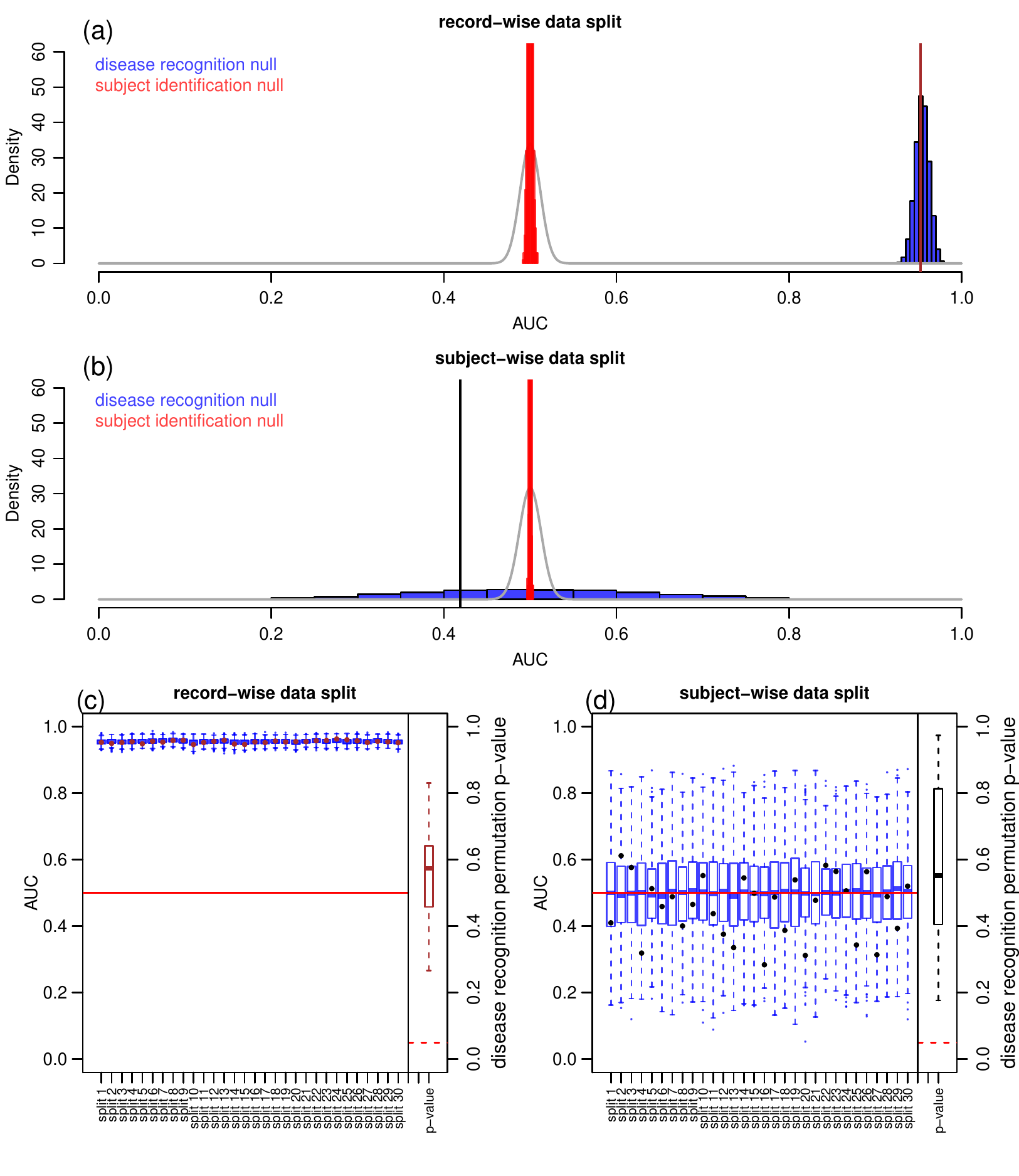}
\caption{Voice data example. Panel a shows the disease recognition permutation null (blue) the observed AUC value (brown line), the identity confounding permutation null distribution (red), and the density of the analytical null distribution used for the computation of the pseudo p-values, for the record-wise data split. Panel b shows the analogous objects for the subject-wise data split. Panel c reports the disease recognition null distributions (blue boxplots) for 30 distinct record-wise data splits, as well as, the observed AUC values (brown dots), and the distribution of the disease recognition permutation p-values across the 30 data splits (brown boxplot). Panel d shows the analogous objects for the subject-wise data split. The disease recognition nulls was generated with 10,000 permutations in panels a and b, and 1,000 permutations in panels c and d. The identity confounding nulls were generated with 1,000 feature permutations (and 300 label permutations per feature permutation).}
\label{fig:mPowerVoiceExample}
\end{center}
\end{figure}

Figure \ref{fig:mPowerVoiceExample} shows the results for the voice data. From panel a we observe a massive amount of identity confounding for the record-wise split data, with the disease recognition null distribution (blue) concentrated at very high AUC values (median AUC $= 0.954$). Both the identity confounding permutation p-value and the pseudo p-value are 0 in this example. The disease recognition p-value ($= 0.5843$), on the other hand, shows that the random forest classifier is not performing disease recognition, even though the observed AUC (brown line) is $0.952$. Panel b reports the results for the subject-wise data split, and show that, once again, the classifier is not learning about the disease labels (disease recognition p-value $= 0.7243$).

In order to assess the robustness of these results with respect to the training/test data split, we show in panel c the disease recognition null distributions (blue boxplots) for 30 distinct record-wise data splits, as well as, the respective AUC values (brown dots), and the distribution of the disease recognition permutation p-values across the 30 data splits (brown boxplot). The results show that in none of the 30 data splits we observed a permutation p-value smaller than 0.05 (dashed red line), showing that the very high AUC values (brown dots) achieved by the classifier are due largely to the classifier's ability to identify subjects. Finally, panel d reports the results for the subject-wise data split, confirming that the classifier is indeed unable to perform disease recognition.

\begin{figure}[!h]
\begin{center}
\includegraphics[width=\linewidth]{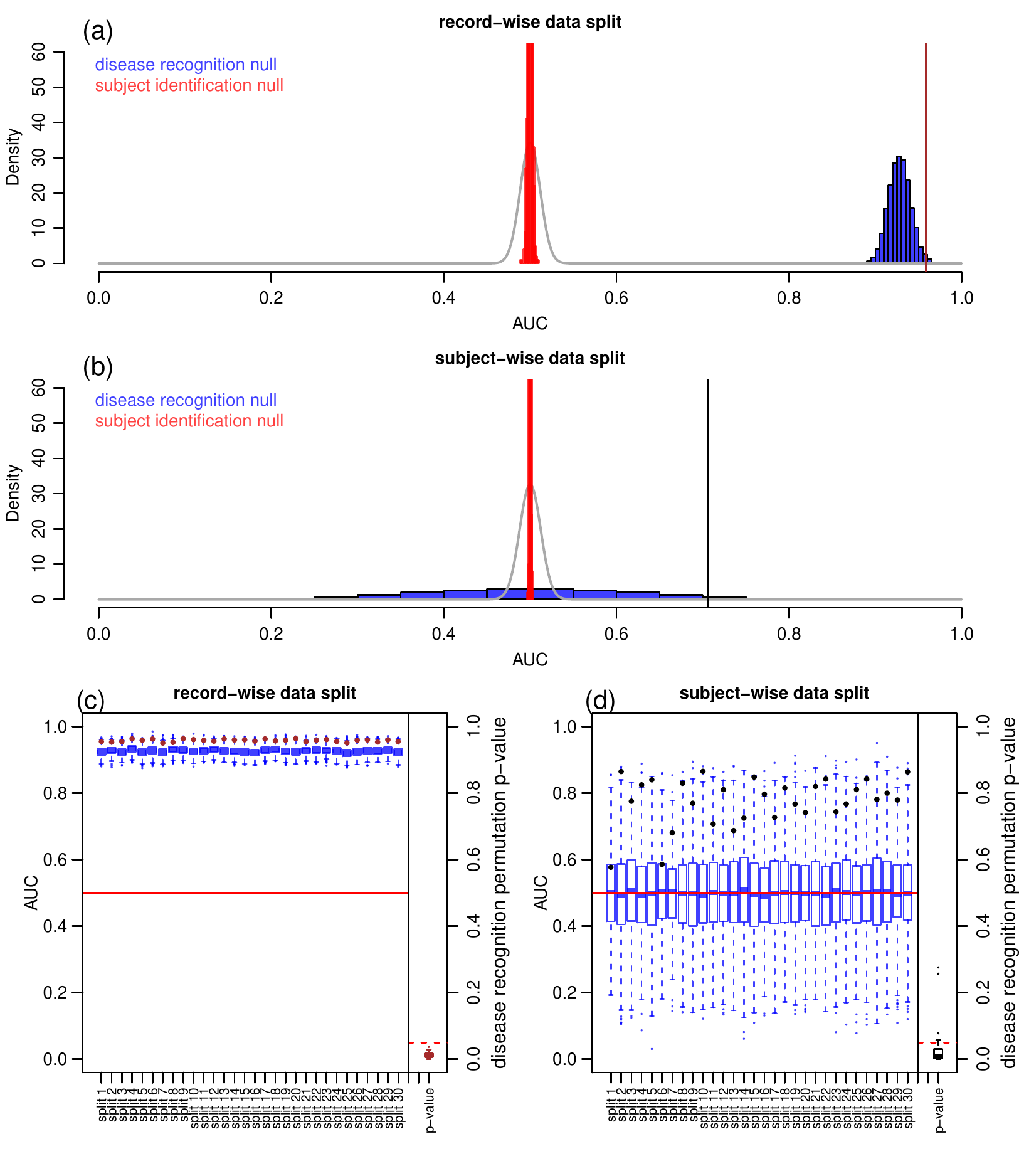}
\caption{Tapping data example. Panel a shows the disease recognition permutation null (blue) the observed AUC value (brown line), the identity confounding permutation null distribution (red), and the density of the analytical null distribution used for the computation of the pseudo p-values, for the record-wise data split. Panel b shows the analogous objects for the subject-wise data split. Panel c reports the disease recognition null distributions (blue boxplots) for 30 distinct record-wise data splits, as well as, the observed AUC values (brown dots), and the distribution of the disease recognition permutation p-values across the 30 data splits (black boxplot). Panel d shows the analogous objects for the subject-wise data split. The disease recognition nulls was generated with 10,000 permutations in panels a and b, and 1,000 permutations in panels c and d. The identity confounding nulls were generated with 1,000 feature permutations (and 300 label permutations per feature permutation).}
\label{fig:mPowerTappingExample}
\end{center}
\end{figure}

\begin{figure}[!h]
\begin{center}
\includegraphics[width=\linewidth]{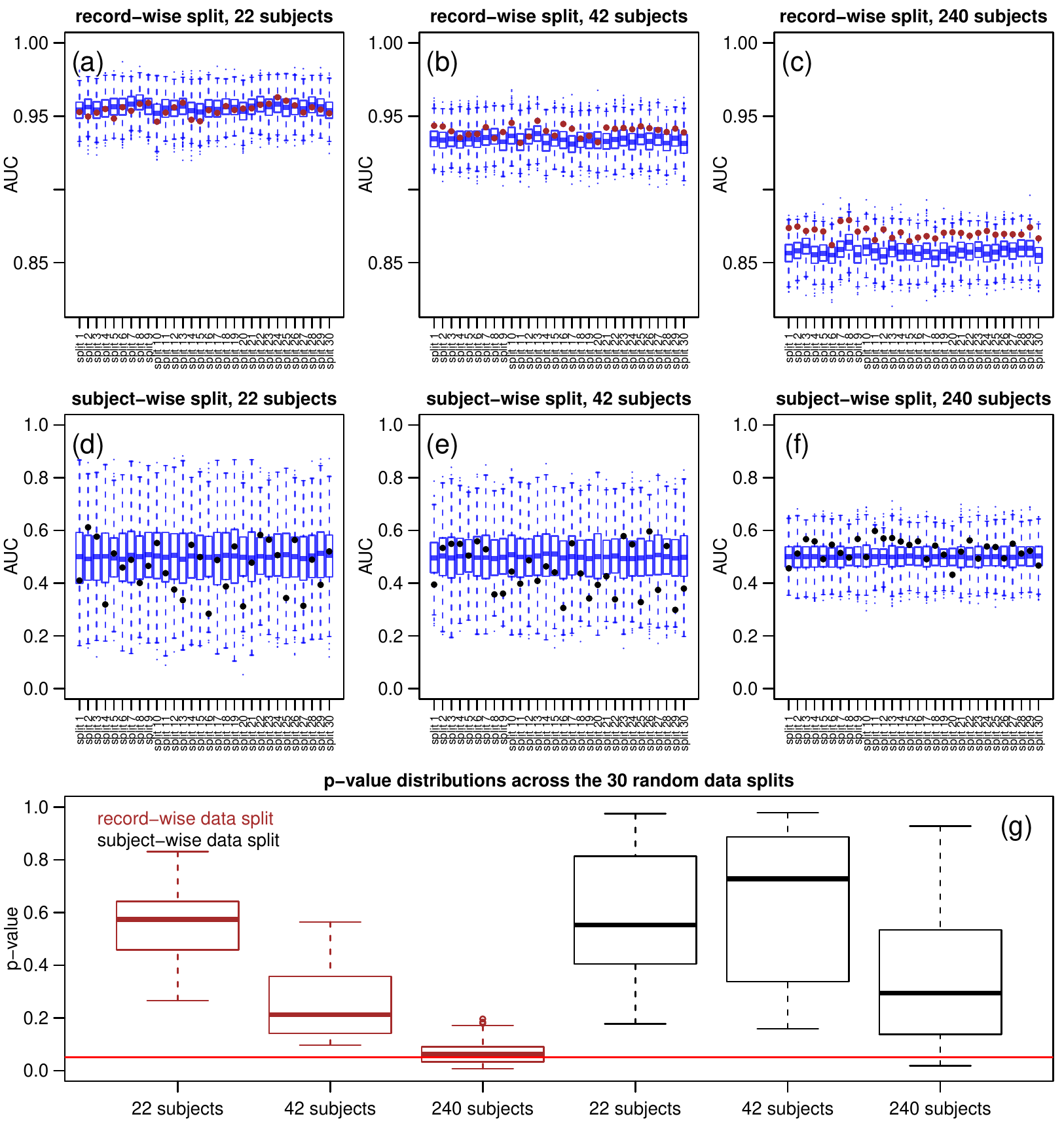}
\caption{Comparison of the disease recognition performance in the voice data, across increasing number of subjects. Panels a, b, and c, show the disease recognition null distributions (blue boxplots), across 30 distinct record-wise data splits, for classifiers trained and evaluated using, respectively, 22 subjects (with at least 100 records per subject), 42 subjects (with at least 50 records per subject), and 240 subjects (with at least 10 records). The brown dots represent the observed AUC values. Observe the decrease in AUC, as the number of subjects increase. Panels d, e, and f, show the analogous results for 30 distinct subject-wise data splits (with black dots representing the observed AUC values). Note the decrease in the spread of the disease recognition null distributions, as the number of subjects increases. (Observe, as well, the different y-axis scales of panels a, b, and c, when compared to panels d, e, and f.) Panel g shows the distributions of the disease recognition permutation p-values, across the 30 random data splits shown in panels a to f. Note that in none of the data splits the permutation p-value was smaller than 0.05 (red line), showing that the classifier is unable to perform disease recognition with the adopted voice features.}
\label{fig:mPowerVoiceIncreasingNSubjects}
\end{center}
\end{figure}

\begin{figure}[!h]
\begin{center}
\includegraphics[width=\linewidth]{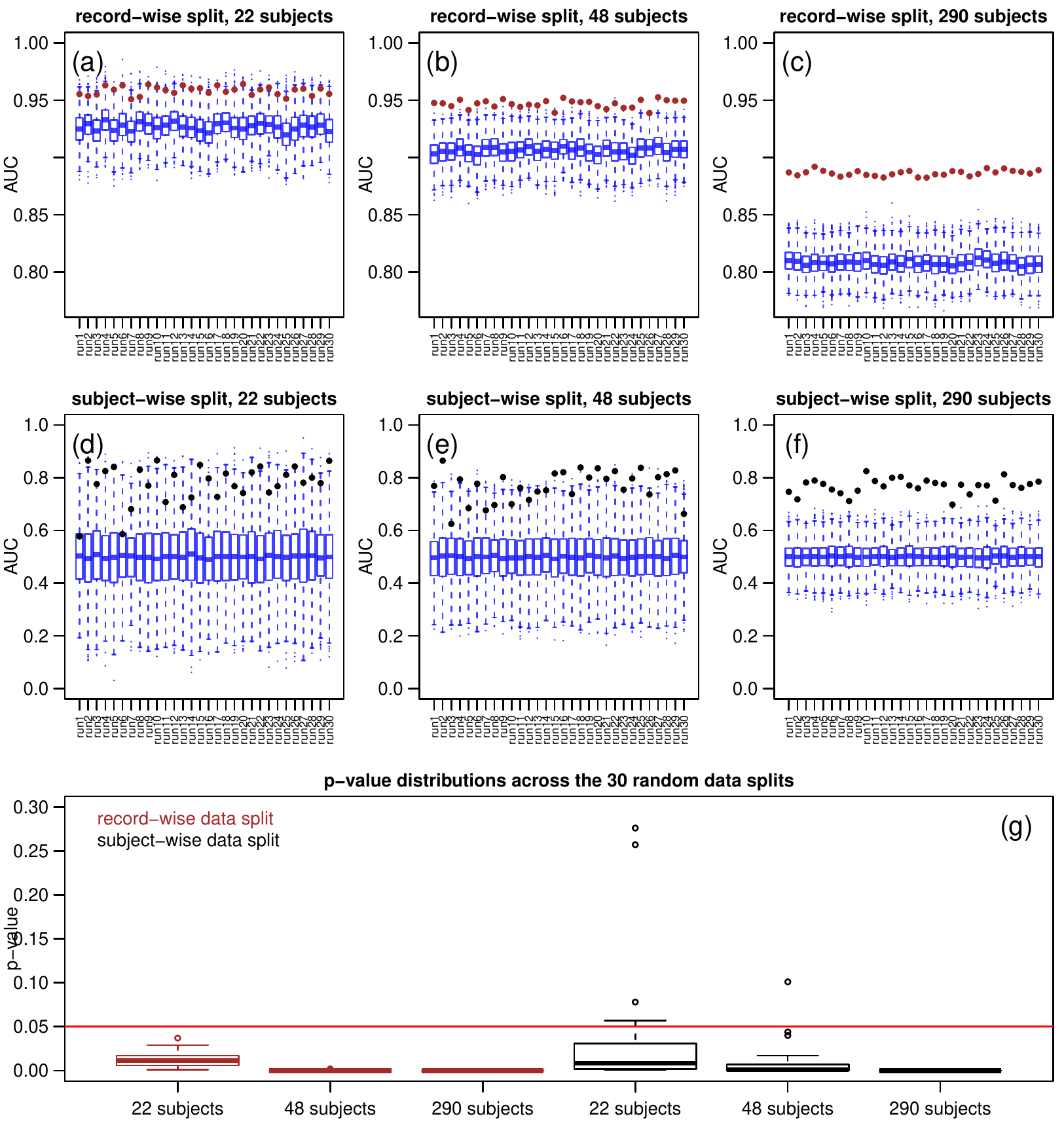}
\caption{Comparison of the disease recognition performance in the tapping data, across increasing number of subjects. Panels a, b, and c, show the disease recognition null distributions, across 30 distinct record-wise data splits, for classifiers built with, 22 subjects (with at least 100 records per subject), 48 subjects (with at least 50 records per subject), and 290 subjects (with at least 10 records), respectively. The brown dots represent the observed AUC values. Observe the decrease in AUC, as the number of subjects increase. Panels d, e, and f, show the analogous results for 30 distinct subject-wise data splits (with black dots representing the observed AUC values). Note the decrease in the spread of the disease recognition null distributions, as the number of subjects increases. Panel g shows the distributions of the disease recognition permutation p-values, across the 30 random data splits shown in panels a to f. Note how the p-values tend to decrease, as the number of subjects increase. In particular, the accentuated improvement in power for the subject-wise split strategy (black boxplots), illustrates how the increase in the number of subjects reduced to a great extent the model under-fitting issue observed in panel d.}
\label{fig:mPowerTappingIncreasingNSubjects}
\end{center}
\end{figure}

Figure \ref{fig:mPowerTappingExample} shows the results for the tapping data. Again from panel a we observe high levels of identity confounding for the record-wise data split (identity confounding permutation p-value and the pseudo p-value are again 0), but contrary to the voice data, the random forest classifier is now able to perform disease recognition with the tapping features (disease recognition p-value $= 0.014$). Panel b, shows that the classifier trained with the subject-wise data split is still able to perform disease recognition, although the statistical significance is weaker (permutation p-value $= 0.055$). Panel c shows that the results presented in panel a are robust across distinct record-wise data splits. Panel d assesses the robustness of the findings observed in panel b, and shows that results tend to be more variable and less significant in this case when compared to the record-wise data splits (note how the distribution of the disease recognition p-values have a larger spread than in the record-wise case). We point out, nonetheless, that this larger variability is likely due to the very small number of subjects and the heterogeneous nature of the data, and suggests that a certain amount of model under-fitting is going on in the tapping data.

In order to assess whether a larger number of subjects would improve the power to perform disease recognition (especially when adopting the subject-wise data split strategy) we repeated our analysis using a larger number of subjects (who, nonetheless, provided a smaller number of records). Specifically, for the voice data we analyzed two additional (age matched and male) cohorts containing 42 subjects (who provided at least 50 records), and 240 subjects (who contributed at least 10 records), whereas for the tapping data we analyzed two additional cohorts containing 48 and 290 subjects (who contributed at least 50 and 10 records, respectively). As before, we ran the analyses on 30 distinct data splits, with half of the subjects and (approximately) half of the records assigned to training set in the subject-wise and record-wise data splits, respectively.

Figure \ref{fig:mPowerVoiceIncreasingNSubjects} reports the results for the voice data. Panels a, b, and c show the disease recognition null distributions and observed AUC values for the record-wise data splits across increasing numbers of subjects. In all 3 panels, the disease recognition null distributions are concentrated around very high AUC values, suggesting that identity confounding is playing an important role in all three cohorts. It is interesting to note that the AUC values tended to decrease, as the amount of data used to train the random forests increased. This drop in AUC is, nevertheless, not surprising, given that the classifier is mostly performing subject identification and that a larger number of subjects makes it harder to identify the individual subjects. Observe, as well, that the random forest only started to detect the disease signal in panel c, where a larger amount of data was available to train the classifier (note how the brown dots started to approach the upper tail of the blue boxplots). Panels d, e, and f, show analogous results for the subject-wise data splits. Note that, as expected, the disease recognition null distributions were always centered around 0.5, and their spread tended to decrease as the number of subjects increased. Observe, nonetheless, that the random forest was unable to perform disease recognition even when the training set contained data from 120 subjects (panel f). Panel g shows the distributions of the disease recognition permutation p-values, across the 30 random data splits shown in panels a to f.

Figure \ref{fig:mPowerTappingIncreasingNSubjects} shows the results for the tapping task. Overall, we observe the same patterns as for the voice data, except that tapping features seem to be less prone to perform subject identification (comparison of panels a, b, and c on Figures \ref{fig:mPowerTappingIncreasingNSubjects} and \ref{fig:mPowerVoiceIncreasingNSubjects} show that both the observed AUC values and disease recognition null distributions tend to be located at lower AUC values for the tapping data than for the voice data), and are better able to perform disease recognition in both record-wise and subject-wise data splits (compare Figures \ref{fig:mPowerTappingIncreasingNSubjects}g and \ref{fig:mPowerVoiceIncreasingNSubjects}g).

\section{Discussion}

In this paper, we propose the use of permutation tests to detect identity confounding in clinical machine learning applications. By focusing on non-parametric hypothesis tests, instead of the more difficult task of generalization error estimation via cross-validation, we are able to statistically test if the classifier is performing disease recognition and/or subject identification, even in situations where the cross-validation error estimates are unreliable due to assumption violations\cite{little2017}.

We illustrate the application of the proposed tests with synthetic data, as well as, with real data from a Parkinson's disease study\cite{bot2016,trister2016}. While our illustrations are all based on the AUC metric, these permutation tests can be implemented with other performance metrics. Also, even though this paper focuses primarily on diagnostic applications, our permutation tests can be applied in other binary classification problems (in which case the ``disease recognition" null distribution should be just renamed as the ``label recognition" null). Furthermore, these tests can also be readily applied to multi-class problems, by simply adopting a multi-class performance metric as a test statistic for the label recognition null, and the median of the label recognition null as the test statistic for the identity confounding null. (Note that the subject-wise label shuffling used to generate the label recognition null is essentially the same as in the binary case, except that instead of the two colors representing the case and control labels in Figures \ref{fig:recsplitsubjshuffle} and \ref{fig:subsplitsubjshuffle}, we would have as many colors as the number of classes.)

The main drawback of the proposed identity confounding permutation test is the large amount of computation required to generate the null distribution (recall that the computation of the test statistic requires $p_l$ subject-wise label permutations, and we need to generate this statistic $p$ times, leading to a total of $p \times p_l$ computations). However, as illustrated in Supplementary Figure S2, it is possible to adopt a smaller number of subject-wise label permutations, than of record-wise feature permutations, without sacrificing the accuracy of the results. Still, another strategy to reduce computation is to first compute the ``pseudo p-value" described in Section 2.3, to see if it is even necessary to compute the permutation p-value. (As a matter of fact, the pseudo p-values alone, would have been enough to establish the presence of identity confounding in our analyses of the mPower data.) Observe, however, that the pseudo p-value shortcut is only available for the AUC metric, and cannot be used with other binary classification or multi-class performance metrics.

For illustration purposes, we applied the identity confounding permutation test to both record-wise and subject-wise data splits. In practice, however, it is not necessary to test for identity confounding when adopting subject-wise data splits, as the influence of any digital fingerprints are automatically neutralized by the subject-wise strategy. Hence, in practice, we recommend the following steps to determine which data split strategy is the most sensible to assess classification performance in the data-set at hand. Starting with record-wise data splits, run the disease recognition permutation test in order to informally assess if identity confounding is playing a role. Next, test for the presence of identity confounding using the pseudo p-value. If identity confounding is detected, then the subject-wise data split should be used. Otherwise, if the pseudo p-value does not support the presence of identity confounding, or if a metric other than the AUC is being used, run the computationally expensive identity confounding permutation test. If identity confounding is not detected, then the record-wise data split can be safely used to assess the classifier performance. Otherwise, run the disease recognition permutation test using the subject-wise data split in order to assess classification performance.

In addition to the data exchangeability assumption (required by any permutation test) our tests also implicitly assume that the classifier is properly tuned, so that an inability to perform disease recognition or subject identification is really due to the lack of signal in the data, and not poor algorithm tuning. Note that in our analyses, we adopt the random forest classifier, which requires minimal amounts of tuning, while still achieving similar classification performance to other classifiers that require much more careful tuning.

Ours tests are designed to detect identity confounding only. Of course, other sources of confounding (e.g., age and sex) can still play a role in the classification performance, and need to be dealt with before running the permutation tests. In the mPower study, the age distributions of cases and controls are not well balanced (with the control population being younger than the PD cases population). Hence, in our analyses, we adopted a subset of age matched case and control participants. Furthermore, because very few female participants contributed enough records, we restricted our analyses to male participants only. (We point out, that had we not matched the subjects by age, we would not be able to tell whether the disease recognition permutation test was performing disease recognition or ``age recognition" of young versus old participants, or still both disease and age recognition. In any case, the identity confounding permutation test would still be able to detect the presence of digital fingerprints since age would be confounded with the disease labels, but not with the subject identities.)

Our synthetic data examples clearly illustrate that identity confounding may arise: (i) when the feature distributions have similar shapes across the subjects, but show serial dependencies across the records within each subject (as illustrated in Figure \ref{fig:example1}b); as well as, (ii) when the data shows differences in location and/or spread, but no serial association across the records (as illustrated in Figures \ref{fig:example4}b and \ref{fig:example5}b). Furthermore, we were only able to avoid identity confounding with data simulated under the very special setting where the feature data was independent and identically distributed within and across all subjects (and we could not conceive of a data generation mechanism that could be used to simulate data consistent with both the alternative hypothesis for disease recognition and the null hypothesis for identity confounding). Note, nonetheless, that we do not make any claims that the particular model adopted to simulate the feature data is able to closely reproduce the dependency/distributional structures observed in real mobile data (as any simulated example will always be artificial to some extent). Rather, the key point we wanted to communicate is that many different sources of heterogeneity in the feature data can easily give rise to identity confounding (and that our permutation tests are able to detect it). The important practical implication, is that identity confounding might turn out to be the rule, rather than the exception, in digital health studies, since it seems reasonable to expect that feature data from real applications will likely show some amount of heterogeneity across subjects.

In particular, our analyses of the voice and tapping data, collected by the mPower study, showed undeniable evidence of a high degree of identity confounding for classifiers built using record-wise data splits. Although it is true that our results also showed evidence of model under-fitting for classifiers trained with subject-wise data splits, we point out that model under-fitting can be ameliorated by simply increasing the number of subjects used in the analyses. (The rationale is that, as the number of subjects used to train the classifier increases, the chance that the training set is missing a critical part of the pattern that relates features to disease labels decreases, so that the classifier has a better chance to generalize to new unseen cases, even when the data is fairly heterogeneous). But, most importantly, our data does not support the hypothesis raised in Little \textit{et al}\cite{little2017} that model under-fitting (alone), rather than identity confounding, could explain the discrepancy in classification performance between the record-wise and subject-wise strategies.

Interestingly, the voice data seems to require a much larger number of subjects in order to perform disease recognition. The fact that the random forest classifier was unable to perform disease recognition with the subject-wise data splits, and only started to detect the disease signal with record-wise data splits when the training set contained over 100 subjects, combined with the fact that the voice-based classifier achieved higher AUC values than the tapping-based one in the record-wise data splits, suggests that our voice features are more vulnerable to identity confounding and less suited to perform disease recognition, when compared to the tapping features. A possible explanation, is that voice features are intrinsically better able to capture personal and physiological characteristic of a subject compared to a game based tapping feature. An alternative explanation, is that the weaker disease recognition signal might be due to suboptimal processing and cleaning of the voice raw data. To see why, note that the digital fingerprint of a subject may arise not only from biological characteristics, such as the pitch of a subject's voice, but also from non-biological/environmental artifacts, such as the amount of background noise and the distance of the subject's mouth to the phone's microphone. These non-biological artifacts can certainly contribute to identify subjects when the voice tasks are performed under different conditions across the subjects, but under consistent conditions within each subject. (Note that while this issue can be somewhat controlled in laboratory experiments, this is not generally the case in mobile health studies, such as mPower, that are conducted at home under uncontrolled conditions.) This problem is likely exacerbated by the use of highly sensitive sensors, capable of detecting quite subtle variations in sound. The tapping task, on the other hand, seems to provide less room for environmental artifacts.

Any source of identity confounding, be it a biological signal or an environmental artifact, is automatically neutralized by the subject-wise data splitting strategy. It is certainly true that the adoption of the subject-wise data split in applications based on multiple records from a small number of subjects can lead to model under-fitting. However, it seems natural that a classifier trained on a small number of subjects, won't generally be able to achieve a stellar classification accuracy, and that one should not expect to achieve state-of-the-art disease recognition performance using only a handful of case and control subjects. Furthermore, as already discussed above, model under-fitting can be reduced by increasing the number of subjects used to train the classifier.

In view of all the points discussed above, we expect that identity confounding will be a common occurrence in diagnostic digital health applications, and that, in general, the subject-wise split will be the most sensible data splitting strategy for diagnosis (in agreement with the recommendations in Saeb \textit{et al} \cite{saeb2017}, Sarkar \textit{et al}\cite{sarkar2010,sarkar2013} and Chaibub Neto \textit{et al}\cite{chaibubneto2017}). In any case, we would still suggest, as a best practice, to always test for the presence of identity confounding in record-wise data splits (if only to confirm its existence, and empirically justify the adoption of the subject-wise splitting). An interesting follow up study (that is out of the scope of the present paper, and is left as future work) would be to apply the proposed permutation tests to several publicly available mobile health data-sets, and empirically assess the frequency and strength of the identity confounding issue in digital health studies.

\section*{Acknowledgements}

This work has been founded by the Robert Wood Johnson Foundation.

\clearpage

\beginsupplement

\begin{figure}[!h]
\begin{center}
\includegraphics[width=\linewidth]{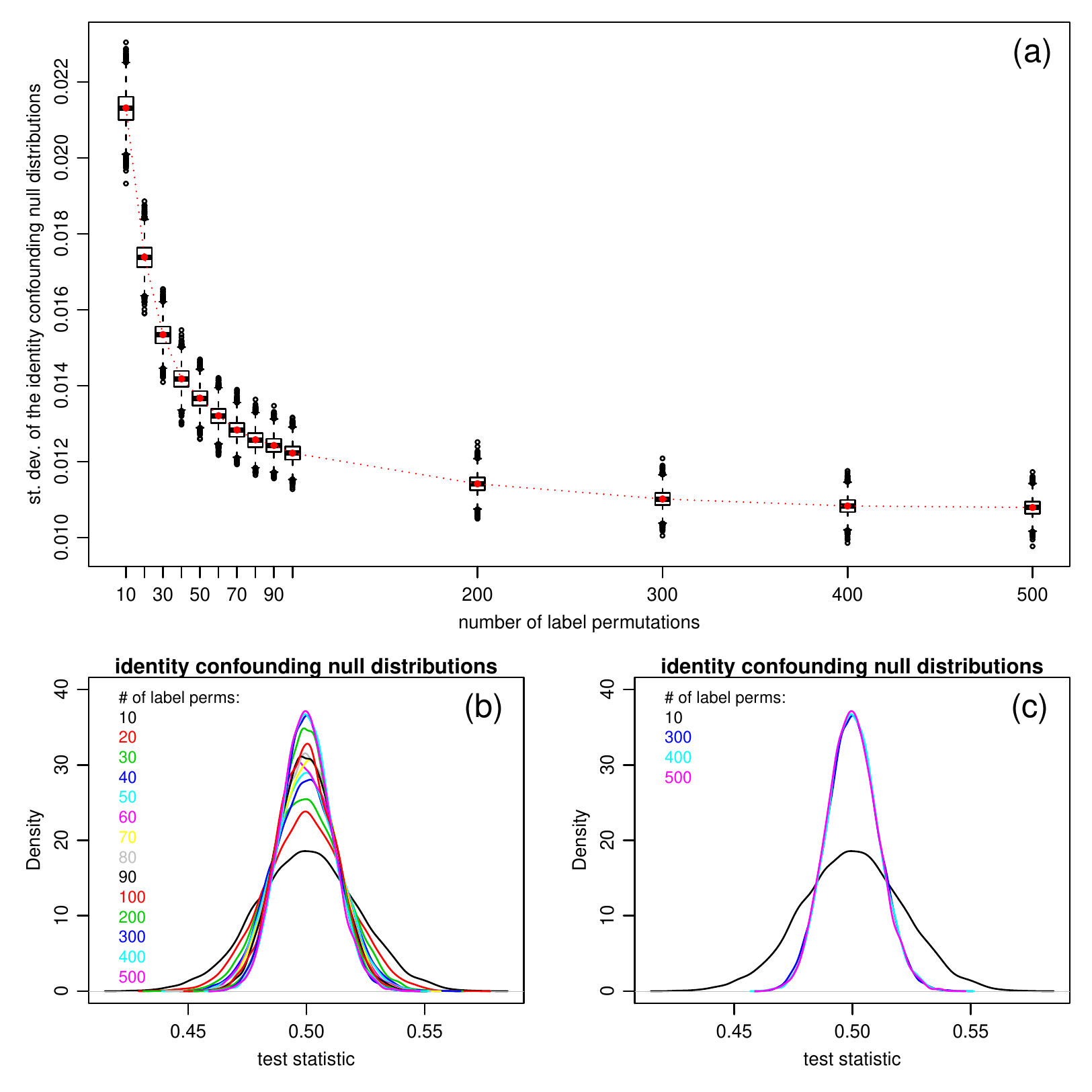}
\caption{Determination of a sufficient number of subject-wise label permutations in the generation of the identity confounding null distribution. As described in Algorithm 2, in the main text, the generation of the identity confounding null distribution requires the computation of the disease recognition null distribution (which is based on $p_l$ subject-wise label shufflings), for each one of the $p$ record-wise feature data permutations, so that the total number of computations is given by $p \times p_l$. Note, nonetheless, that because the test statistic ($\tilde{m}^{\ast}$) corresponds to the median of a performance metric ($m^\ast$), we don't need to use a large number of label permutations, $p_l$ (since it is reasonable to expect that the median of the disease recognition null distribution, computed with, say, 1,000 subject-wise label permutations should be close to the median computed with 10,000 permutations). This rationale suggests, that we might be able to determine a sufficient number of label permutations, by computing the identity confounding null distribution using increasing $p_l$ values, and inspecting when the null distribution ``stabilizes". Panel a shows the standard deviation of the identity confounding null distribution (red dots) computed with $p = 10,000$ record-wise feature permutations for $p_l$ varying from 10 to 500 subject-wise label permutations. The black boxplots show the distributions of the standard deviations across 10,000 sub-samples of size 1,000 of the respective identity confounding null distributions, and provide estimates of the amount of sampling variability we should expect when we adopt 1,000 record-wise feature permutations, rather than 10,000. The panel shows that, while the standard deviation estimates based on $p = 10,000$ permutations (red dots) are still monotonically decreasing as $p_l$ increases, the sampling variability that we see when we adopt $p = 1,000$ by far overwhelms the small difference in standard deviation that is achieved by increasing $p_l$ from 300 to 400 (or 500). Similarly, panel b shows that the identity confounding null distribution tends to stabilize as the number of label permutations increases, and panel c shows that the null distributions are already very similar when we adopt $p_l = \{300, 400, 500\}$. All the results were based on data simulated according to synthetic example 1 (equation \ref{eq:example1}), in the main text.}
\label{fig:suppleFig1}
\end{center}
\end{figure}

\begin{figure}[!h]
\begin{center}
\includegraphics[width=\linewidth]{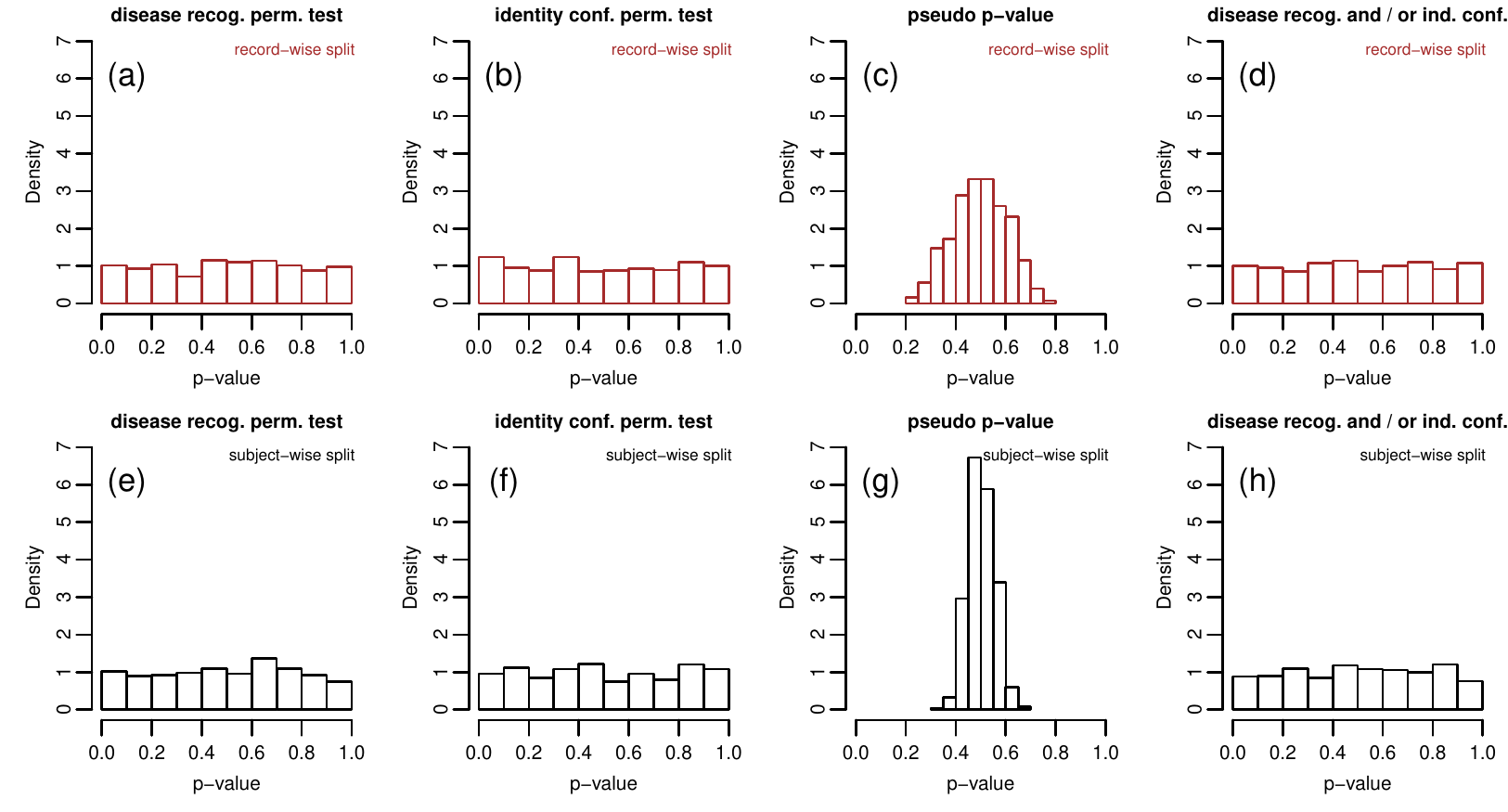}
\caption{Assessing type I error rate control with a simulation study. We simulated 500 distinct data-sets under the null hypothesis of both the disease recognition and the identity confounding tests. For each synthetic data-set, the feature data of each subject was independently simulated according to the model, $\bfX_s = c \, \bfV_s + d \, \bfE_s$, in order to allow for correlation across the features (see main text for further details). Each simulated data-set was generated with a unique combination of simulation parameters, with the scalars $c$ and $d$ varying from 0.1 to 2, the number of cases and controls varying from 5 to 10, and the number of records per participant varying from 10 to 20. In order to select parameter values spread as uniformly as possible over the entire parameter ranges, we adopted a Latin hypercube space filling design in the determination of the parameter values used to generate each data-set. Due to computational constraints, the permutation tests were based on 100 permutations. The top panels show the results from tests applied to data-sets split in a the record-wise fashion, while the bottom panels report the results based on subject-wise data splits. Panels a and e show the p-value distributions from the disease recognition permutation test, while panels b and f show the p-value distributions from the identity confounding permutation test. As expected, the distributions are approximately uniform showing that the type I error rate of the permutation tests are being controlled at the nominal significance levels. Panels c and g show the distributions of the pseudo p-values. As expected, the distributions are bell-shaped since a central limit theorem for the sample median of a random variable is kicking in (note that because the pseudo p-value is computed using the median of the AUC metric, and the tail probability computed with the median of a sample of test statistics is equal to the median of the tail probabilities computed for each test statistic in the sample, that is, $1 - \Phi((\mbox{median}_j\{\mbox{auc}_j^\ast\} - 0.5)/\phi) = \mbox{median}_j\{1 - \Phi((\mbox{auc}_j^\ast - 0.5)/\phi)\}$, we have that the pseudo p-value distributions actually correspond to the distributions of the sample median of disease recognition p-value distributions). Finally, panels d and h, report the p-value distributions from the analytical test for the presence of disease recognition and/or identity confounding ($H_0^{\ast\ast\ast}$) calculated according to expression (\ref{eq:null3pval}), in the main text. As expected, the p-value distributions are also approximately uniform.}
\label{fig:suppleFig2}
\end{center}
\end{figure}

\end{document}